\documentclass[journal]{IEEEtran}
\usepackage{amsmath}
\usepackage{hhline}
\usepackage{multirow}

\DeclareMathOperator*{\argmin}{arg\,min}

\ifCLASSINFOpdf
	\usepackage[pdftex]{graphicx}
\else
	\usepackage[dvips]{graphicx}
\fi

\usepackage{cite}
\usepackage{url}
\usepackage[ruled,vlined]{algorithm2e}
\usepackage{mathtools}
\usepackage{subcaption}

\usepackage{tikz}
\usepackage{pgfplots}
\usepackage{cisco}

\usepackage{multirow}
\usepackage[normalem]{ulem}
\useunder{\uline}{\ul}{}

\begin{document}
\title{Maintaining an Up-to-date Global Network View in SDN}
\author{Mohamed~Aslan,
        and~Ashraf~Matrawy,~\IEEEmembership{Senior~Member,~IEEE}
	\thanks{
		M. Aslan is with the Department of Systems and Computer Engineering, Carleton University, Ottawa, ON K1S 5B6, Canada.
		e-mail: maslan@sce.carleton.ca
	}
	\thanks{
		A. Matrawy is with the School of Information Technology, Carleton University, Ottawa, ON K1S 5B6, Canada.
		e-mail: ashraf.matrawy@carleton.ca
	}
	\thanks{
		This paper extends a preliminary version published in the IEEE Communication Letters vol. 20 (1) 5-8 \cite{aslan2015impact}.
	}
}


\maketitle

\begin{abstract}
Maintaining an up-to-date global network view is of crucial importance for intelligent SDN applications that need to act autonomously.
In this paper, we focus on two key factors that can affect the controllers' global network view and subsequently impact the application performance.
Particularly we examine: (1) network state collection, and (2) network state distribution.
First, we compare the impact of active and passive OpenFlow network state collection methods on an SDN load-balancing application running at the controller using key performance indicators that we define. We do this comparison through: (i) a simulation of a mathematical model we derive for the SDN load-balancer, and (ii) an evaluation of a load-balancing application running on top of single and distributed controllers.
Further, we investigate the impact of network state collection on a state-distribution-aware load-balancing application.
Finally, we study the impact of network scale on applications requiring an up-to-date global network view in the presence of the aforementioned key factors.
Our results show that both the network state collection and network state distribution can have an impact on the SDN application performance.
The more the information at the controllers becomes outdated, the higher the impact would be.
Even with those applications that were designed to mitigate the issues of controller state distribution, their performance was affected by the network state collection.
Lastly, the results suggest that the impact of network state collection on application performance becomes more apparent as the number of distributed controllers increases.

\end{abstract}


\IEEEpeerreviewmaketitle

\section{Introduction}

\IEEEPARstart{T}{raditionally}, computer networks consist of a number of switches, routers, middleboxes (such as firewalls, load-balancers and intrusion detection systems) and/or other hosts. In most cases, network administrators would have to manually configure those devices either by accessing the device through a console port, or via a remote access protocol. The configuration process was a hard one as network administrators need to be aware of how to configure each of these devices individually, especially if they are manufactured by different vendors \cite{vaughan2011openflow}. In some cases, they fail to inter-operate with each other as some of those devices might be running a proprietary protocol that can only inter-operate with similar devices made by the same vendor. Even in the case of standard and open protocols, oftentimes different implementations behave differently. Kuzniar \textit{et al.} \cite{kuzniar2012soft} created an approach for testing the inter-operability of OpenFlow-enabled \cite{mckeown2008openflow} switches.

On the other hand, Software-Defined Networking (SDN) is a promising network architecture that proposes the decoupling of data (also forwarding) and control planes.
It employs a logically centralized controller powered by a network-wide global view to orchestrate the network \cite{fundation2012software}.
It enables innovation by transforming the task of network administration to network programming.
In principle, SDN attempts to overcome the manageability problems of legacy networks, by transforming statically configured networks to dynamically programmed ones. In SDN, network applications ought to be logically installed at the physically separated control plane known as the ``controller''. Those applications communicate with the switches in order to provide the desired network functionality \emph{e.g.,} Network address translation (NAT), and load-balancing.

However, with this great shift in how networks are perceived, many new network application design challenges have appeared. Those challenges include, but not necessarily limited to, collecting up-to-date network state information by the controllers in order to maintain the global network view, and using multiple controllers in large scale networks. Such challenges, if not correctly handled, could affect the performance, security or scalability of the network, leading to higher Operating Expenditure (OPEX), and hence SDN application developers need to be ware of their presence.

The Open Networking Foundation (ONF)\footnote{\url{http://www.opennetworking.org}} is a non-profit organization concerned with promoting the concepts of SDN, and with standardizing the OpenFlow \cite{openflowspec15,mckeown2008openflow} protocol. OpenFlow is a southbound API \footnote{A southbound API is a protocol that allows controllers to communicate with the switches. While, northbound APIs enable SDN applications to communicate with SDN controllers.} between the controller and data planes, enabling possible integration of different multi-vendor components, and allowing them to inter-operate. Although there exist other southbound APIs (e.g. ForCES \cite{yang2004forwarding} by the IETF, and Cisco OpFlex \cite{smith2014opflex}), in this paper we only focus on OpenFlow as it is now considered the de-facto standard in SDN-enabled devices.

The Open Networking User Group (ONUG) has recently showed interest in network state collection \cite{onug-correlation}. They proposed an open architecture for designing SDN applications which includes modules for network state collection, network state correlation, and network state analytics. Further, they defined some requirements and usecases for network state collection, correlation and analytics. However, they did not define ``how'' should those modules be developed in order to provide such requirements.

To maintain a global network view, controllers need to gather state information from: (1) the switches, and (2) other controllers (in case of a distributed controller platform). In this paper, we refer to the process of gathering state information from the switches as ``network state collection'', and to the process of exchanging collected state information among the controllers as ``controller state distribution''. We show that, depending on the application requirements, both the network state collection and the controller state distribution mechanisms need be considered as they can impact the controllers' global network view and hence the application performance.

In this paper, we study two main factors that can impact the global network view at the controllers, namely: network state collection, and controller state distribution. In particular, we make the following contributions:
\begin{itemize}
	\item Contribution I:\footnote{This contribution extends our earlier contribution presented in \cite{aslan2015impact} with more traffic models.} We study the impact of passive and active OpenFlow state collection mechanisms on certain key performance indicators that we define. To study this:
	\begin{enumerate}
		\item[a.] we ran simulations based on a mathematical model we derived for the load-balancing (LB) application using different network application performance indicators (\S\ref{sec-model-eval}),
		\item[b.] we developed a load-balancing application and ran experiments in the context of single and distributed controller environments (\S\ref{sec-exp-eval}).
	\end{enumerate}

	\item Contribution II: We investigate the impact of network state collection on controller state distribution aware applications (\S\ref{sec-lsvs}). To study this: we experimented with an implementation of the Least-loaded Server Variation Synchronization (LSVS) distributed load-balancing application (originally proposed by Guo et al. \cite{Guo201495}).
	\item Contribution III: We study the impact of network scale on applications that require an up-to-date network view (\S\ref{sec-large-scale-net}).

For all of the aforementioned contributions, we employed low-variation and high-variation Poisson traffic and On/Off Pareto traffic loads.
\end{itemize}

The rest of this paper is organized as follows.
We provide an overview on the design challenges in developing distributed SDN applications in \S\ref{sec-background}.
\S\ref{sec-related-work} presents related work.
In \S\ref{sec-netview}, we discuss controllers' global network views and challenges faced when maintaining an up-to-date global network view.
\S\ref{sec-prob-statement} presents the problem statement.
In \S\ref{sec-model-eval}, we evaluate a LB application model. We show our experimental setup and evaluation for the LB application in \S\ref{sec-exp-eval}.
The impact of network state collection on state distribution aware applications is shown in \S\ref{sec-lsvs}.
In \S\ref{sec-large-scale-net}, we show the impact of controllers' state distribution on the SDN application performance when the number of controllers increases.
Finally \S\ref{sec-conclusion} will be our conclusion and an outline for possible foreseeable work.

\section{Background on SDN Controllers}
\label{sec-background}

The use of a physically centralized controller can limit the scalability and reliability of large scale networks. A single controller can represent a bottleneck as well as a single point of failure. Recent research in SDN \cite{dixit2013towards, tootoonchian2010hyperflow, levin2012logically, koponen2010onix, berde2014onos} as well as the SDN architecture defined by the ONF, highlight the use of logically centralized but physically distributed controllers. Contrary to a centralized controller, distributed controllers can scale-out \footnote{By \emph{scale-out} we mean the ability of network applications to scale by increasing the number of controllers.} by installing new controllers, and can be fault-tolerant in case of controller failure. Designing SDN applications that run on top of distributed controllers is a non-trivial task due to the complexity of handling controllers state synchronization which in-turn can affect the applications' performance. For the rest of the paper, we refer to a physically centralized controller as a ``single controller'', while we call logically centralized but physically distributed controllers ``distributed controllers''.

Brewer theorem \cite{brewer2000towards, brewer2012cap}, also known as CAP theorem, states that it is impossible for a distributed system to simultaneously provide the following guarantees: Consistency, Availability, and Partition tolerance, and that there is always a trade-off between the system's consistency and availability in the presence of network partitions.
For example, in the case of a datastore cluster that is comprised of a number of distributed nodes. Assume that at one point in time, those nodes got partitioned due to a network failure, and new data update requests continued to arrive at some of the cluster's nodes. If those nodes continued to handle the requests, the stored data might become inconsistent and the cluster will still be available (\emph{i.e.,} it handled the requests). Otherwise, the data would remain consistent but the cluster would not be available (\emph{i.e.,} requests failed).
As the CAP theorem applies to any distributed system, it would imply that designing SDN applications running on top of distributed controllers encounters a trade-off between consistency and availability, in case of network partitions \cite{panda2013cap}.

\section{Related Work}
\label{sec-related-work}

There exist a multitude of projects that investigate the feasibility of using multiple SDN controllers. In this section, we survey some of those projects.

In DevoFlow \cite{curtis2011devoflow} project, the authors observed the overhead of switches frequently invoking the controller and concluded that it is insufficient for flow setup in high-performance networks. They suggested that a distributed controller might be able to handle such higher loads. The authors presented DevoFlow which addresses the issues of efficient control and statistics collection in OpenFlow.

The authors of the FlowVisor \cite{sherwood2009flowvisor} proejct brought the concept of hypervisor-like network virtualization into SDN. It is an OpenFlow proxy sitting between the control and data planes, which acts as a network virtualization layer that lies between the network devices and control applications, allowing the use of multiple \emph{guest} OpenFlow controllers at the same time (one controller per slice) in order to ensure that slices are isolated from one another.

HyperFlow \cite{tootoonchian2010hyperflow} is a distributed controller platform that is built around publish/subscribe messaging \cite{tarkoma2012publish} and event-based paradigms.
It is implemented on-top of a flexible distributed file system, where applications have control over consistency and durability. HyperFlow was designed with resiliency in-mind so that in the presence of network partitioning, partitions can continue to operate independently, favoring availability.

Onix \cite{koponen2010onix} provides an API for implementing applications that run on-top of distributed SDN controllers. The authors of Onix, realized that applications often have different requirements for the consistency of the network state they manage. Hence, Onix gives applications the flexibility to make their own trade-offs between consistency and scalability. This trade-off flexibility is achieved by providing the control applications with two storage subsystems that employ two different consistency models: (1) a strongly consistent transactional Database (DB), and (2) an eventually consistent memory-based single-hop Distributed Hash Table (DHT).

Open Network Operating System (ONOS) \cite{berde2014onos} employs instances of Floodlight \cite{floodlight} controllers, as well as a distributed DB to build a distributed Network Operating System (NOS) platform for SDNs. The authors tested two prototypes of ONOS. The first prototype uses Cassandra \cite{lakshman2009cassandra, lakshman2010cassandra}, an eventually consistent distributed DB to maintain the global network view and to provide scalability and fault-tolerance. On top of that, the first prototype includes a graph DB and provides an API for SDN applications to query the graph DB. The evaluation of the first prototype showed that it suffered from low performance due to the complexity of its data model, the excessive data store operations, and the controllers had to keep polling network state information from the switches. The second prototype was developed with the objective of improving the low performance issues of the first one. The Cassandra DB was replaced with a distributed in-memory key-value data store with a lower latency, and a caching layer for the network topology information was deployed. Moreover, a publish/subscribe messaging sub-system was incorporated for event notification.
Even though ONOS was developed after ONIX, it did not provide the SDN application developers the flexibility to choose between either a strong or an eventual consistency model.

DISCO (DIstributed SDN COntrol plane) \cite{phemius2014disco} employs instances of Floodlight \cite{floodlight} SDN controllers and uses an AMQP-based publish/subscribe message bus, in order to provide a distributed control platform for multi-domain SDN networks. DISCO supports software agents that can be dynamically installed at different controllers at run-time in order to add support for required functionalities \emph{e.g.,} QoS.

\section{Global Network View}
\label{sec-netview}

As aforesaid, we are interested in the impact of the recency of controllers' global network view on network application performance as well as the factors that could affect that global network view. In this section, we review and discuss two of such factors: (1) network state collection, and (2) controller state distribution.

\subsection{Network State Collection}
There exist different mechanisms that network application developers can employ in collecting the network state information required for maintaining an up-to-date network view.
Excluding \emph{middle-boxes} which are likely to face scalability issues, we classify these techniques into: (1) non-OpenFlow-based, and (2) OpenFlow-based.
The former include the use of other protocols as SNMP \cite{case1989simple} or sFlow \cite{phaal2001inmon}.
Teh latter involve the use of information stored at the controllers as well as \emph{flow} or \emph{port} statistics that are polled from the switches.
In this paper we are only interested in OpenFlow-based mechanisms.

Curtis \textit{et al.} \cite{curtis2011devoflow} stated that OpenFlow maintains different statistics counters such as: packets, bytes and flow duration, and provides the controller with different approaches for collecting those statistics from the switches. They classified those approaches into: (1) push-based, and (2) pull-based. The push-based approaches are in the form of notifications sent to the controller when a specific event occurs, \emph{e.g.,} new flow arrives or a flow is removed. The pull-based approaches are in the form of responses to messages sent by the controller to retrieve the counters that the switches maintain.

We divide OpenFlow-based mechanisms into: (1) \emph{passive}, and (2) \emph{active} based on whether the controller injects new messages to collect the state information or not.
The choice of whether to design a network application to base its decisions on passive, active, or hybrid state collection could be a design challenge.

\subsubsection{Passive Network State Collection}
When a new packet arrives at a switch with no flow rule to match, by default, the packet's header will be forwarded to the controller. The controller is then responsible for deciding what to do with the packet and for instructing the switches.
The controller can locally keep track of flow rules inserted into the switches, which we call \emph{passive} state collection.
However since by default not every packet is forwarded to the controller, the controller-maintained information are mainly flow-based information and may not include packet or byte information.
Yu et al. \cite{yu2013flowsense} proposed a passive network monitoring approach that computes the utilization of the links between switches, discussing how their approach can be combined with \emph{active} approaches.

\subsubsection{Active Network State Collection}
Controllers can asynchronously communicate with the switches they control in order to request the required state information.
An example is the flow and port statistics.
OpenFlow defines the format of the messages that can be exchanged by the controllers and the switches.
With regards to applications that require up-to-date state information, they need to periodically communicate with the network switches as such information could be outdated relative to the application needs.
It is also important to highlight that our results (\S VII-B) indicate that the period at which the controllers poll the necessary information from the switches can impact the performance of those applications.

\subsection{Controller State Distribution}
One major challenge is how to keep the controllers' network views consistent. With distributed controllers, as each controller collects state information about the network, they need to exchange their views in order to build a global network view. The consistency between controllers is governed by the consistency model employed \cite{panda2013cap}.

To study the impact of inconsistent network views at the controllers on SDN application performance, Levin et al. \cite{levin2012logically} created a simple distributed load-balancing application based on periodic synchronization (controllers are synchronized every fixed time period). They used a flow simulator to study the design of distributed SDN applications.
First, they evaluated the impact of inconsistent global network view on the performance of a LB application.
They showed that the inconsistency can significantly degrade the performance of SDN applications.
Second, they investigated the application complexity versus robustness against outdated states. They concluded that applications that are aware of the underlying state distribution can avoid depending solely on outdated states when making decisions, and perform better than those that are unaware.
However, their implementation suffered from two issues, identified by Guo et al. \cite{Guo201495}.
First, in order to achieve an acceptable load-balancing performance, short synchronization periods are required, causing high synchronization overhead.
Second, their LB was vulnerable to forwarding loops due to the inconsistency of network views at the controllers.

Guo et al. \cite{Guo201495} further explored and mitigated the aforesaid issues by introducing a new state distribution mechanism that overcame the issues of controllers synchronization overhead, and also fixed the issue of forwarding loops.
To solve the first issue, Guo et al. proposed a new controller synchronization mechanism called Load Variance-based Synchronization (LVS). LVS is a trigger-based synchronization mechanism that achieves low synchronization overhead by conducting controller state synchronizations \emph{only} when the load of a specific server exceeds a certain threshold. LSV exploited the fact that in load-balancers only the information about the least-loaded server needs to be synchronized and not the load of each server, resulting in a low synchronization overhead. They presented two LB designs based on LVS: (1) Least loaded Server Variation Synchronization (LSVS), and (2) Least loaded Domain Variation Synchronization (LDVS). The only difference is that the latter is based on the load variation of a whole domain instead of a single server.

In order to solve the the second issue, Guo et al. had to ensure that all the controller agree on the least-loaded server (even if the information is incorrect). Thus, they introduced controller roles, where a controller can be either an active controller or passive controller. An active controller (only one controller can be active at a time) is the controller that manages the domain of the least-loaded server. The active controller is also responsible for instantiating the synchronization process when the load on the least-loaded server exceeds a certain threshold. Finally, the performance of the LVS LB was dependent on the value of that threshold. It is worth mentioning that since LVS exploits the fact that only the information about the least-loaded server needs to be synchronized, it may not be suitable for the use in other SDN applications.

\section{Problem Statement}
\label{sec-prob-statement}

In SDN, maintaining an up-to-date global view at the controllers is crucial for the performance and security of the network. This global view requires that controllers collect network state information regularly from the switches.

We study how the recency of the network state information collected by the controllers, and the consistency among controllers, affect the global network view and subsequently the network application performance.
In particular, we study the impact of network state collection on the performance of SDN applications. In case of distributed SDN controllers, we also study the impact of controller state distribution on the performance of SDN applications.

In order to achieve that, we design a load-balancing SDN application that can run on top of a single controller or distributed controllers environments. Then, we use this LB to study the impact of network state collection and controller state distribution on the SDN application performance.

\subsection{The \emph{Relative Difference} as a Performance Indicator}
In order to study the application performance, performance indicators must first be defined. Performance indicators are application-specific metrics that are used to evaluate the application performance. Hence network application developers need to carefully choose these indicators.

We study the behavior of a LB application (see \S\ref{lb-design} for design) when employing different mechanisms for network state collection. For simplicity, we consider the load-balancing between two servers. We opted for the relative difference between the traffic \emph{received} by each server as an indicator for the LB's performance. The smaller the difference, the better the performance. The relative difference can be defined in terms of flows ($\xi_{f}$) or bytes ($\xi_{b}$).
We chose the relative difference as a normalized indicator (unit-less) to be able to compare ($\xi_{f}$) and ($\xi_{b}$) where needed.

Equations \ref{eqn:xi-flow} and \ref{eqn:xi-byte} show the relative difference in terms of flows and bytes, respectively. In case of $\xi_{f}$, $f_{i}$ represents the flow count (as measured at the server) assigned to server $i$ (in our experiments, $i \in \{1, 2\}$) at a given time.
While for $\xi_{b}$, $b_{i}$ represents the amount of traffic received (in bytes, measured at the server) by server $i$ at a given time.

\begin{equation}
	\small
	\xi_{f} = \frac{| f_{1} - f_{2} |}{f_{1} + f_{2}} \quad \cdots \quad s.t. \quad 0 \le \xi_{f} \le 1
	\label{eqn:xi-flow}
\end{equation}

\begin{equation}
	\small
	\xi_{b} = \frac{| b_{1} - b_{2} |}{b_{1} + b_{2}} \quad \cdots \quad s.t. \quad 0 \le \xi_{b} \le 1
	\label{eqn:xi-byte}
\end{equation}

The relative difference ($\xi$) between the amount of traffic received by each server can work only in the case of two controllers. Later in \S\ref{sec-large-scale-net} we study the performance at a larger scale network with more than two controllers. In such case, we use the standard deviation ($\sigma$) of the amount of traffic received by each server (shown in (\ref{xload-std})) as the performance indicator.

\subsection{Load-Balancer's Design}
\label{lb-design}
A simplified version of the load-balancing algorithm is presented in Algorithm \ref{lbalgo}.
When a new client's request (packet) arrives at a switch and there are no rules in the switch's flow-table on how to process this request, the switch will forward the request to its controller.
The controller will, according to its local view of the network, decide where to assign the flow associated with the request.
In case of distributed controllers, a controller can assign the flows to its local domain server or forward the flows to the out-of-domain server connected to the other switch.
The two controllers periodically synchronize their state. On each synchronization period, the controllers exchange their local view of the network.
We use a hard-time of 2 sec for all flows, \emph{i.e.,} a flow rule lives in the switch's flow-table for only 2 sec, then it has to be reassigned.

We consider four different variation for a LB: (1) a single-controller LB that uses passive state collection (SP),
(2) a single-controller LB that uses active state collection (SA), (3) a two-controller distributed LB that uses passive state collection (DP),
and (4) a two-controller distributed LB that uses active state collection (DA).
The objective of these LBs is to reduce the difference between the servers' link utilization, hence reduce $\xi$ ($0$ is the optimum).
Table \ref{tbl-exp-summary} shows summary of the experiments we ran.

\begin{algorithm}
	\small
	\DontPrintSemicolon
	\SetKw{KwGoTo}{goto}
	\KwData{$S_{n}$, set of n servers.}
	\KwData{$L_{n}$, set of traffic load of the n servers (could be measured in flows or bytes)}
	\Begin{
		\While{$pkt$ arrives}{
			$l_{min} \leftarrow \infty$\;
			\ForEach{$s \in S_{n}$}{
				\If{$L(s) < l_{min}$}{
					$l_{min} \leftarrow L(s)$\;
					$s_{min} \leftarrow s$\;
				}
			}
			SetupPath($pkt$, $s_{min}$)\;
		}
	}
	\caption{SDN load-balancing at the controllers.}
	\label{lbalgo}
\end{algorithm}

\section{Model Evaluation}
\label{sec-model-eval}
In this section, we show how the recency of the controllers' global network view can affect the network application performance (contribution (1)) using a simulation for a LB model that we developed which employs the relative difference ($\xi$) as a performance indicator (see \S\ref{sec-prob-statement}-B).

We derive $\xi_{f}$ only in the case of SP. However, similar steps can be used to derive $\xi_{b}$ for SA.
We make the following assumptions:
(1) the number of switches and servers in the network $N = 2$,
(2) the load-balancing algorithm is invoked for every flow arrival event (as in the case of OpenFlow), and
(3) we do not take network delays into consideration.
Next, we use the following notations:
\begin{itemize}
	\item $E^{i}_{k}$ --- the number of expired flows assigned to server $k$ ($1 \le k \le N$) at a given inter-arrival event $i$.
	\item $L^{i}_{k}$ --- the load on server $k$, at the $i^{th}$ flow inter-arrival event.
	\item $\Delta ^{i}$ --- the difference between the two servers' loads at the $i^{th}$ inter-arrival event.
	\begin{equation}
		\Delta ^{i} = L^{i}_{1} - L^{i}_{2}
		\label{load-diff}
	\end{equation}
	\item $T^{i}$ --- the total loads on the two servers at the $i^{th}$ inter-arrival event.
	\begin{equation}
		T^{i} = L^{i}_{1} + L^{i}_{2}
		\label{load-sum}
	\end{equation}
	\item $\xi_{f} ^{i}$ --- the relative difference between the servers at the $i^{th}$ inter-arrival event.
	\begin{equation}
		\xi_{f}^{i} = |\Delta ^{i}| / T^{i}
		\label{delta-flow}
	\end{equation}
	\item $M^{i}_{k}$ --- number of new flows that will be assigned to server $k$, based on the second assumption $\sum_{k=1}^{N} M^{i}_{k} = 1$.
	\item $d^{i}$ --- the decision parameter of controller.
\end{itemize}
\begin{align}
	d^{i+1} &= (L^{i}_{1} - E^{i+1}_{1}) - (L^{i}_{2} - E^{i+1}_{2})&&\nonumber\\
			&= \Delta ^{i} - E^{i+1}_{1} + E^{i+1}_{2}&&(using (\ref{load-diff}))
\end{align}
\begin{align}
	L^{i+1}_{k} &= L^{i}_{k} - E^{i+1}_{k} + M^{i+1}_{k}
	\label{new-load}
\end{align}
\begin{align}
M^{i+1}_{1} =
\begin{cases}
	0, & \text{if } d^{i} > 0 \\
	1, & \text{if } d^{i} \le 0
\end{cases},\quad 
M^{i+1}_{2} =
\begin{cases}
	0, & \text{if } d^{i} \le 0 \\
	1, & \text{if } d^{i} > 0
\end{cases}
\end{align}
\begin{align*}
	\Delta ^{i+1} &= L^{i+1}_{1} - L^{i+1}_{2}&&(using (\ref{load-diff}))\nonumber\\
	 			&= (L^{i}_{1} \!{-}\! E^{i+1}_{1} \!{+}\! M^{i+1}_{1}) - (L^{i}_{2} \!{-}\! E^{i+1}_{2} \!{+}\! M^{i+1}_{2})&&(using (\ref{new-load}))\nonumber\\
				&= \Delta ^{i} + (M^{i+1}_{1} \!{-}\! M^{i+1}_{2}) + (E^{i+1}_{2} \!{-}\! E^{i+1}_{1})
\end{align*}
\begin{align}
	T^{i+1} &= L^{i+1}_{1} + L^{i+1}_{2}&&(using (\ref{load-sum}))\nonumber\\
				&= L^{i}_{1} \!{-}\! E^{i+1}_{1} \!{+}\! M^{i+1}_{1} + L^{i}_{2} \!{-}\! E^{i+1}_{2} \!{+}\! M^{i+1}_{2}&&(using (\ref{new-load}))\nonumber\\
				&= T^{i} + 1 - (E^{i+1}_{1} + E^{i+1}_{2})&&
\end{align}
\begin{align}
	\xi_{f}^{i+1} &= |\Delta ^{i+1}| / T^{i+1}&&(using (\ref{delta-flow}))\nonumber\\
				&= \frac{|\Delta ^{i} + (M^{i+1}_{1} \!{-}\! M^{i+1}_{2}) + (E^{i+1}_{2} \!{-}\! E^{i+1}_{1})|}{T^{i} + 1 - (E^{i+1}_{1} + E^{i+1}_{2})}&&\IEEEQED
\end{align}

In the cases of DP and DA, each controller maintains its own decision parameter which may differ in values. For DP controllers, they base their decision on values of $L^{i}$ for the local server and $L^{i-s}$ for  the out-of-domain server, where $s$ is a variable that depends on the synchronization period. For DA controllers, they base their decision on values of $L^{i-p}$ for the local server and $L^{i-s}$ for  the out-of-domain server, where $p$ is a variable that depends on the polling period.

\begin{figure}[!t]
\centering
\resizebox{0.5\textwidth}{!}{
	\begin{subfigure}[b]{0.5\textwidth}
		\includegraphics[scale=1.0]{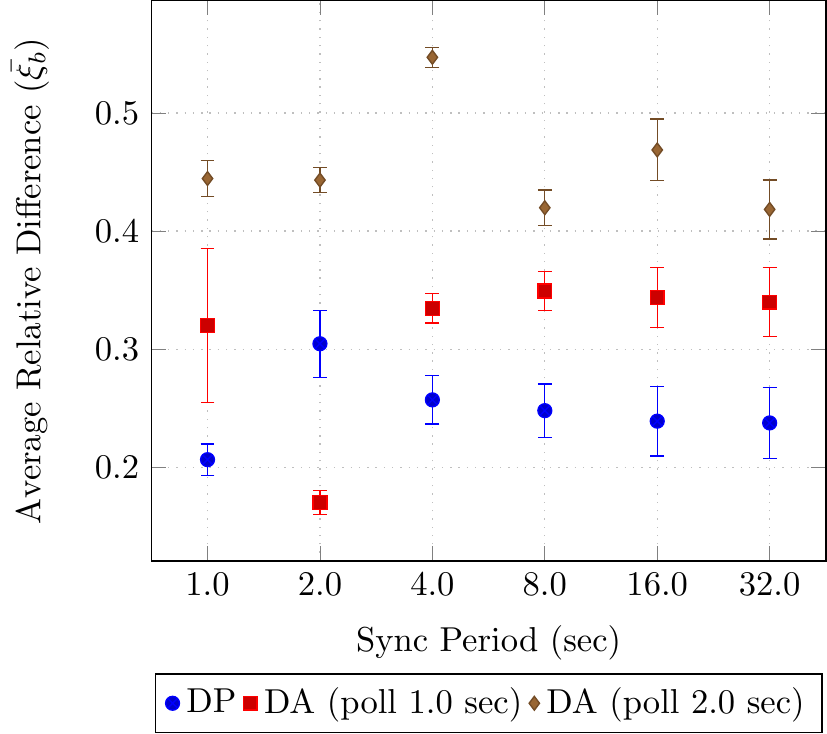}
		\caption{LV}
		\label{fig:F:subfig1}
	\end{subfigure}
	\begin{subfigure}[b]{0.5\textwidth}
		\includegraphics[scale=1.0]{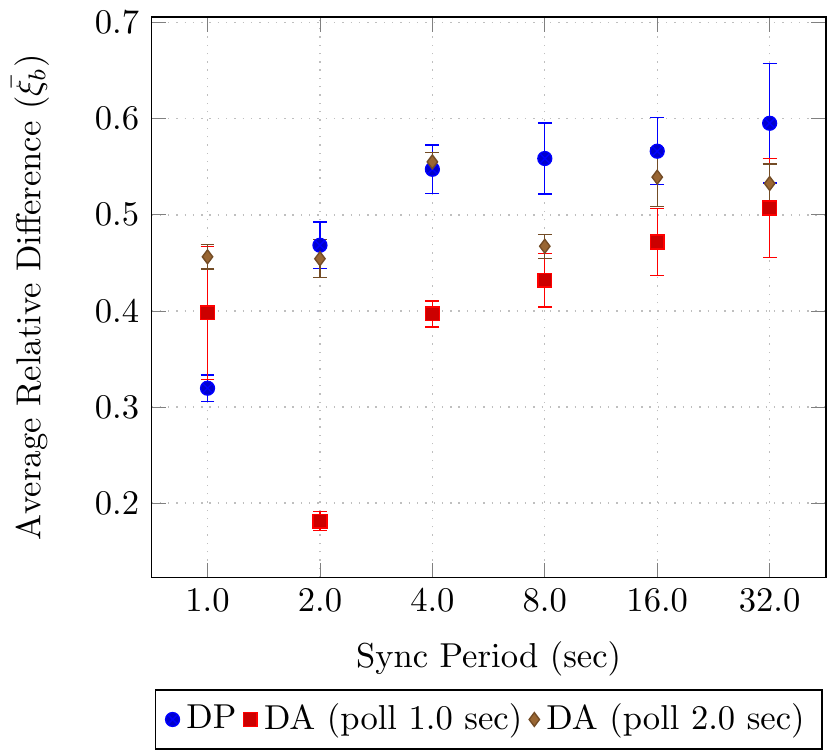}
		\caption{HV}
		\label{fig:F:subfig2}
	\end{subfigure}
}
\caption{Average relative difference ($\bar{\xi_{b}}$) vs synchronization period in case of DP and DA LBs. Results obtained using the model shown in \S VI.}
\label{fig:math-multi-flow-vs-byte}
\end{figure}

Fig. \ref{fig:math-multi-flow-vs-byte} shows the effect of the synchronization period on the performance of both DP and DA.
The results were obtained using the simple traffic load distribution shown Table \ref{traffic-param}.
In the case of LV, DP performed better than the DA (except at sync. period 2). The effect of the polling period had a higher impact than the synchronization on DA.
In the case of HV, DP was more impacted than DA with higher synchronization periods.

\begin{table}[h]
	\caption{
	Employed traffic loads and their parameters in case of the simple traffic load.
	$r_{1}$ and $r_{2}$ are flow arrival rates for switch 1 and 2, respectively.
	$p_{1}$ and $p_{2}$ are packets-per-flow arrival rates for switch 1 and 2, respectively. Payload of any packet is 4KBytes.
	}
	\centering
	\resizebox{.5\textwidth}{!}{
	\begin{tabular}{l | c | c}
	\multicolumn{1}{c}{} & LV & HV\\
	\hhline{-:=:=:}
	Flows & \multicolumn{2}{c}{Poisson process}\\
	Flows rates &  $r_{1}=6 f/s, r{2}=4 f/s$ & $r_{1}=8 f/s, r_{2}=2 f/s$\\
	Flow TTL & \multicolumn{2}{c}{2 sec}\\
	Pkts-per-flow & \multicolumn{2}{c}{Poisson process}\\
	Pkts-per-flow rates & $p_{1}=34 p/s, p_{2}=30 p/s$ & $p_{1}=48 p/s, p_{2}=16 p/s$\\
	\hline
	\end{tabular}
	}
	\label{traffic-param}
\end{table}

\section{Experimental Evaluation}
\label{sec-exp-eval}

In this section, we discuss the setup of our experimentation environment (contribution (1)) which we used for evaluating our different LB implementations (SA, SP, DA and DP).
In our experimentation, we used two different traffic models: (1) a simple Poisson-based traffic load (shown in Table \ref{traffic-param}), and (2) a more realistic taffic load based on an On/Off Pareto process (shown in Table \ref{pareto-traffic-param}). We discuss the results obtained from the first model in \S\ref{exp-simple-model}, and those results obtained from the second one in \S\ref{exp-realistic-model}. 

\subsection{Environment Setup}

To highlight the impact of employing different network state collection mechanisms, we designed a series of experiments we ran on our emulation setup to show how a LB application running at the controller will perform when it takes actions based on both passive collection of flow information and active collection of byte information polled from switches. We do this in the context of both a single and a distributed SDN controllers environments.

In our setup, we use Mininet \cite{lantz2010network} to emulate our network topologies.
We had to deviate from the default configuration, where Mininet runs Open vSwitch \cite{openvswitch} instances in a shared network namespace;
rather we ran separate instances of Open vSwitch in network separeted namespaces. We do this to guarantee the isolation between the switches, and to facilitate the use of different controllers. We also used POX \cite{mccauley2014pox} (an event-based controller  platform and a Python implementation of NOX \cite{gude2008nox}) in-band controllers to control the switches and implement the LBs.

The consistency model employed in our experiment in the cases of distributed LBs (DP and DA) is known as the \emph{delta} consistency model \cite{singla1997temporal}.
The delta consistency model relaxes its data staleness constraints, where a read returns the last value that was updated at most delta time units prior that read operation. In other words, the delta consistency model guarantees that all the controllers converge to a point after delta time units, where they all see the same shared data values if no new updates occurred.
The choice of the right value of the synchronization period is application-specific. Using very small periods might not be feasible due to various network delays or  communication overhead, while very long ones can badly hurt the performance of the application (as we show in \S VII-B). Also the number of synchronization messages exchanged between the controllers increases with the number of controllers.

Figures \ref{nettopo1} and \ref{nettopo2} show the topology used in our experiment. The topology was used in two scenarios:
(1) a single controller was used (Fig. \ref{nettopo1}), and (2) two distributed controllers were used (Fig. \ref{nettopo2}).
Two OpenFlow-enabled switches are connected via a 1000 Mb/s link.
The setup also includes two servers; each is connected to a separate switch via a 100 Mb/s link.
In the second topology, the network is divided into two domains, where each domain consists of: a switch, a server, a number of clients, and a controller that is responsible for controlling that domain.

Finally, 64 clients (32 at each switch) are connected via 100 Mb/s links. The clients will generate UDP requests and create the traffic.
We experimented with two different traffic load distributions. The first a Poisson process based traffic load which we call \emph{simple} traffic load, while the second is a more \emph{realistic} traffic load that is based on an On/Off Pareto distribution \cite{ns2pareto, arnold2015pareto}. In each we employed two different traffic load parameters. The first we call the low-variation traffic load (LV), while the second is the high-variation traffic load (HV).
Table \ref{traffic-param} shows the parameters of the simple traffic load distribution.
Table \ref{pareto-traffic-param} shows the parameters of the more realistic On/Off Pareto traffic load distribution.

\usetikzlibrary{chains}
\usetikzlibrary{decorations.text}
\usetikzlibrary{decorations.pathmorphing}
\usetikzlibrary{scopes}

\begin{figure}[!t]
\centering
\resizebox{.48\textwidth}{!}{
	\begin{tikzpicture}[
		start chain=going right,
		diagram item/.style={
		on chain,
		join
		}
		]
		\node[label=above:Controller, diagram item] (controller) {\controller};
		{ [start branch=A going below left]
			\node[label={[rotate=0]below right:$\kern-1em$Switch$_{1}$}, diagram item] (switch1) {\switch};
			\node[label=below:Server$_{1}$, start branch=3 going below, diagram item] (server1) {\server};
			\node[label=below:Client$_{1}$, start branch=4 going above left, diagram item] (client1) {\client};
			\node[label=below:Client$_{32}$, start branch=5 going below left, diagram item] (client32) {\client};
			\node[start branch=6 going left, on chain] (dots1) {\textbf{\vdots}$\qquad$};
		}
		{ [start branch=B going below right]
			\node[label={[rotate=-0]below left:$\kern+1em$Switch$_{2}$}, diagram item, ,join=with switch1] (switch2) {\switch};
			\node[label=below:Server$_{2}$, start branch=7 going below, diagram item] (server2) {\server};
			\node[label=below:Client$_{33}$, start branch=8 going above right, diagram item] (client33) {\client};
			\node[label=below:Client$_{64}$, start branch=9 going below right, diagram item] (client64) {\client};
			\node[start branch=10 going right, on chain] (dots2) {$\qquad$\textbf{\vdots}};		
		}

		\scriptsize
		\draw[decoration={text along path, text={1000Mbps},text align={center}},decorate]  (switch1) -- (switch2);
		\draw[decoration={text along path, text={100Mbps},text align={center}},decorate]  (client1) -- (switch1);
		\draw[decoration={text along path, text={100Mbps},text align={center}},decorate]  (client32) -- (switch1);
		\draw[decoration={text along path, text={100Mbps},text align={center}},decorate]  (switch2) -- (client33);
		\draw[decoration={text along path, text={100Mbps},text align={center}},decorate]  (switch2) -- (client64);
		\draw[decoration={text along path, text={100Mbps},text align={center}},decorate]  (server1) -- (switch1);
		\draw[decoration={text along path, text={100Mbps},text align={center}},decorate]  (server2) -- (switch2);
		\draw[decoration={text along path, text={100Mbps},text align={center}},decorate]  (switch1) -- (controller);
		\draw[decoration={text along path, text={100Mbps},text align={center}},decorate]  (controller) -- (switch2);
	\end{tikzpicture}
}
\caption{The network topology with a single controller.}
\label{nettopo1}
\end{figure}
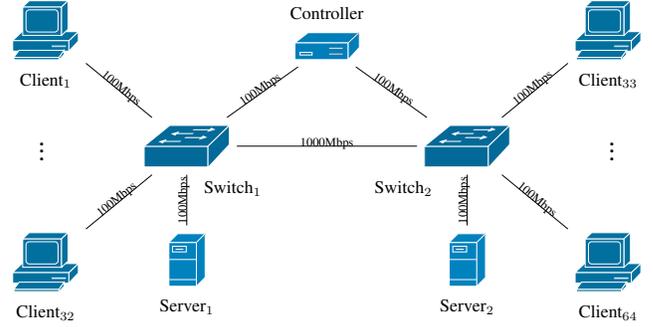

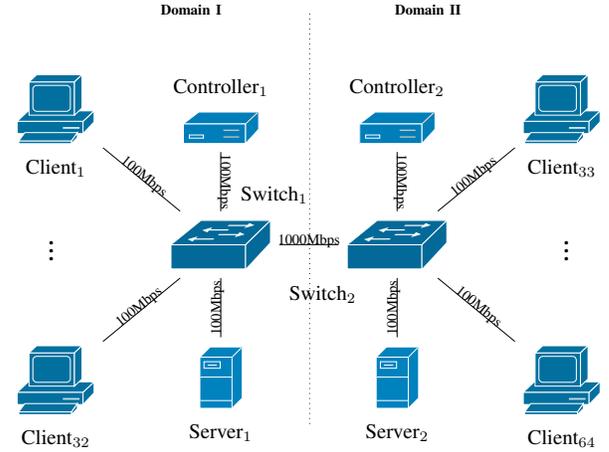
\begin{figure}[!t]
\centering
\resizebox{.45\textwidth}{!}{
\begin{tikzpicture}[
	start chain=going right,
	diagram item/.style={
	on chain,
	join
	}
	]

	\node[label=above:Controller$_{1}$, diagram item, continue chain=going below] (controller1) {\controller};
	\node[label={[rotate=0]above right:$\kern-1em$Switch$_{1}$}, diagram item] (switch1) {\switch};
	\node[label=below:Server$_{1}$, start branch=1 going below, diagram item] (server1) {\server};
	\node[label=below:Client$_{1}$, start branch=2 going above left, diagram item] (client1) {\client};
	\node[label=below:Client$_{32}$, start branch=3 going below left, diagram item] (client32) {\client};
	\node[start branch=4 going left, on chain] (dots1) {\textbf{\vdots}$\qquad$};

	\node[label={[rotate=-0]below left:$\kern+1em$Switch$_{2}$}, diagram item, continue chain=going right] (switch2) {\switch};
	\node[label=above:Controller$_{2}$, start branch=5 going above, diagram item] (controller2) {\controller};
	\node[label=below:Server$_{2}$, start branch=6 going below, diagram item] (server2) {\server};
	\node[label=below:Client$_{33}$, start branch=7 going above right, diagram item] (client33) {\client};
	\node[label=below:Client$_{64}$, start branch=8 going below right, diagram item] (client64) {\client};
	\node[start branch=9 going right, on chain] (dots2) {$\qquad$\textbf{\vdots}};
	
	\scriptsize
	\draw[decoration={text along path, text={1000Mbps},text align={center}},decorate]  (switch1) -- (switch2);
	\draw[decoration={text along path, text={100Mbps},text align={center}},decorate]  (client1) -- (switch1);
	\draw[decoration={text along path, text={100Mbps},text align={center}},decorate]  (client32) -- (switch1);
	\draw[decoration={text along path, text={100Mbps},text align={center}},decorate]  (switch2) -- (client33);
	\draw[decoration={text along path, text={100Mbps},text align={center}},decorate]  (switch2) -- (client64);
	\draw[decoration={text along path, text={100Mbps},text align={center}},decorate]  (server1) -- (switch1);
	\draw[decoration={text along path, text={100Mbps},text align={center}},decorate]  (server2) -- (switch2);
	\draw[decoration={text along path, text={100Mbps},text align={center}},decorate]  (controller1) -- (switch1);
	\draw[decoration={text along path, text={100Mbps},text align={center}},decorate]  (controller2) -- (switch2);
	
	\draw [dotted] (1.5,-5) -- (1.5,2);
	\node at (-0.5,2) {\textbf{Domain I}};
	\node at (3.5,2) {\textbf{Domain II}};
\end{tikzpicture}
}
\caption{The network topology with two distributed controllers.}
\label{nettopo2}
\end{figure}

\begin{figure}[!t]
\centering
\resizebox{0.5\textwidth}{!}{
	\begin{subfigure}[b]{0.5\textwidth}
		\includegraphics[scale=1.0]{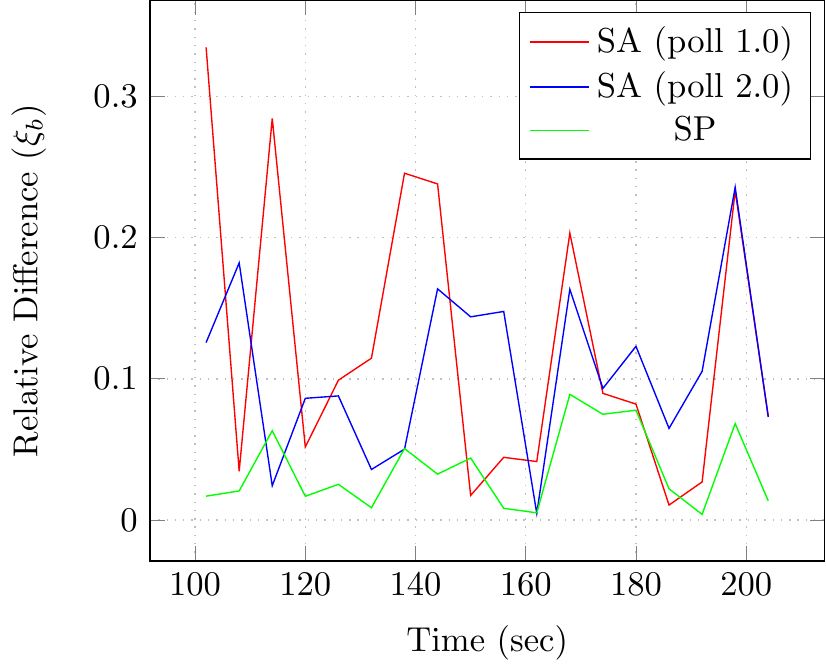}
		\caption{LV}
		\label{fig:A:subfig1}
	\end{subfigure}
	\begin{subfigure}[b]{0.5\textwidth}
		\includegraphics[scale=1.0]{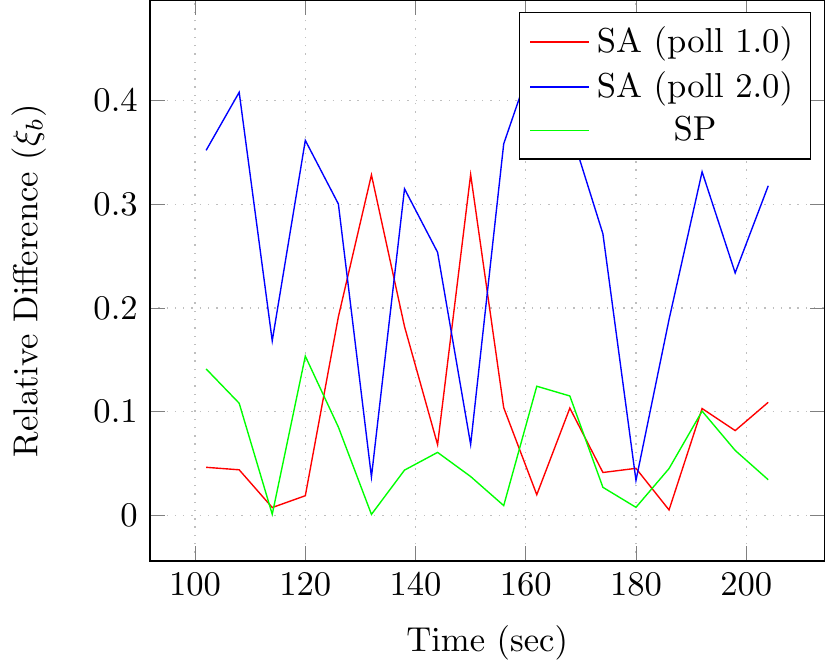}
		\caption{HV}
		\label{fig:A:subfig2}
	\end{subfigure}
}
\caption{Relative difference ($\xi_{b}$) vs time.}
\label{fig:xi-vs-time}
\end{figure}

\begin{figure}[!t]
\centering
\resizebox{0.5\textwidth}{!}{
	\begin{subfigure}[b]{0.5\textwidth}
		\includegraphics[scale=1.0]{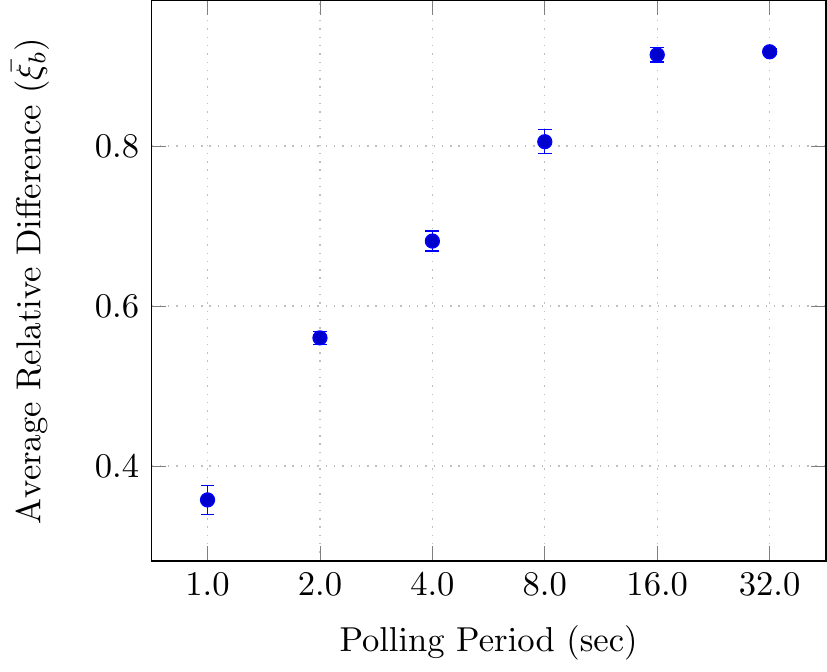}
		\caption{LV}
		\label{fig:B:subfig1}
	\end{subfigure}
	\begin{subfigure}[b]{0.5\textwidth}
		\includegraphics[scale=1.0]{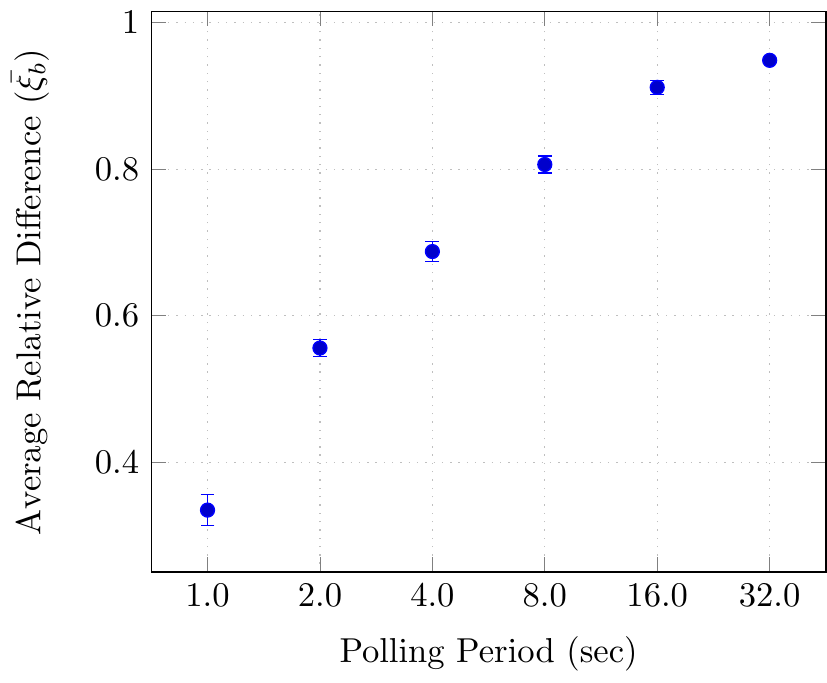}
		\caption{HV}
		\label{fig:B:subfig2}
	\end{subfigure}
}
\caption{Average relative difference ($\bar{\xi_{b}}$) vs polling period in case of a SA LB.}
\label{fig:single-avgxi-vs-poll}
\end{figure}

\begin{figure}[!t]
\centering
\resizebox{0.5\textwidth}{!}{
	\begin{subfigure}[b]{0.5\textwidth}
		\includegraphics[scale=1.0]{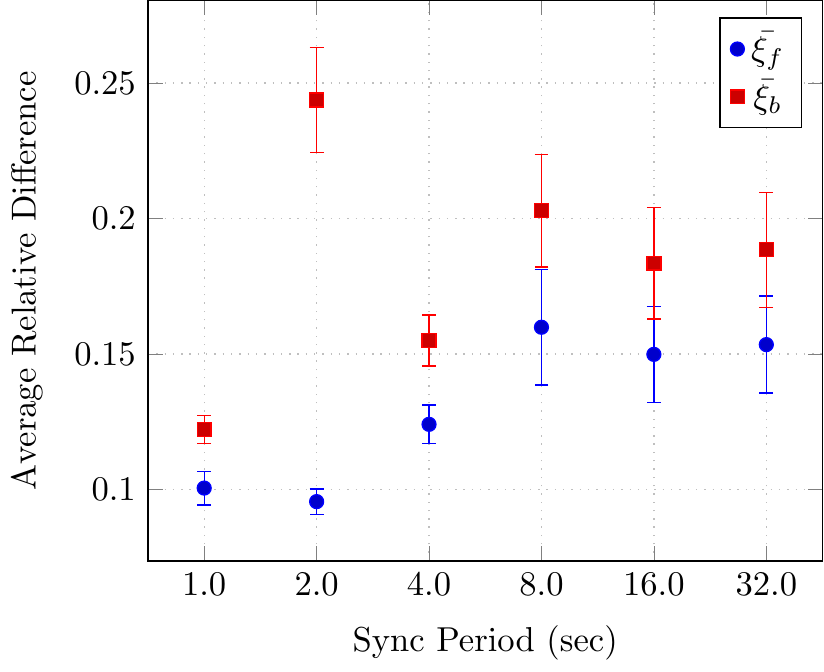}
		\caption{LV}
		\label{fig:C:subfig1}
	\end{subfigure}
	\begin{subfigure}[b]{0.5\textwidth}
		\includegraphics[scale=1.0]{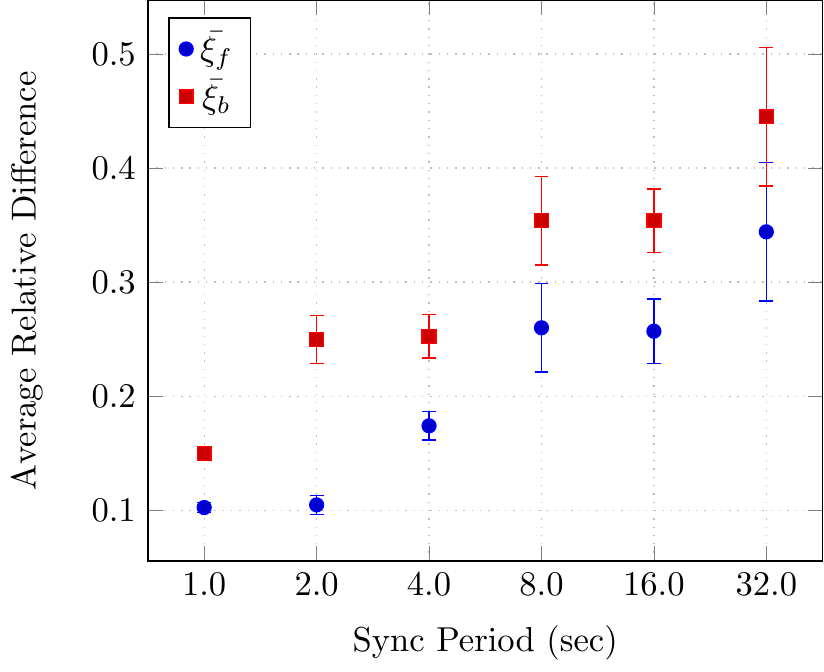}
		\caption{HV}
		\label{fig:C:subfig2}
	\end{subfigure}
}
\caption{Average relative difference in bytes ($\bar{\xi_{b}}$) and flows ($\bar{\xi_{f}}$) vs synchronization period in case of a DP LB.}
\label{fig:multi-avgxi-vs-sync}
\end{figure}

\begin{figure}[!t]
\centering
\resizebox{0.5\textwidth}{!}{
	\begin{subfigure}[b]{0.5\textwidth}
		\includegraphics[scale=1.0]{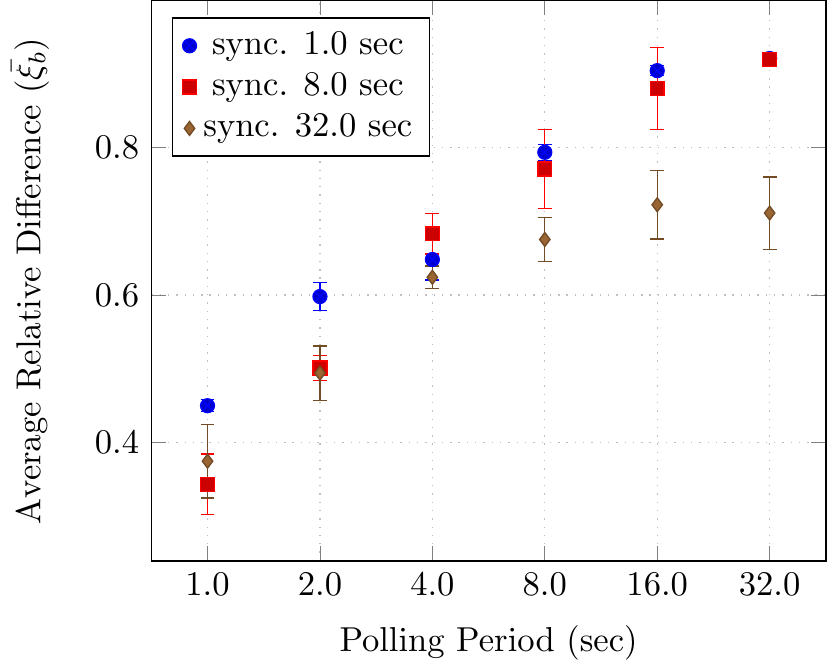}
		\caption{LV}
		\label{fig:D:subfig1}
	\end{subfigure}
	\begin{subfigure}[b]{0.5\textwidth}
		\includegraphics[scale=1.0]{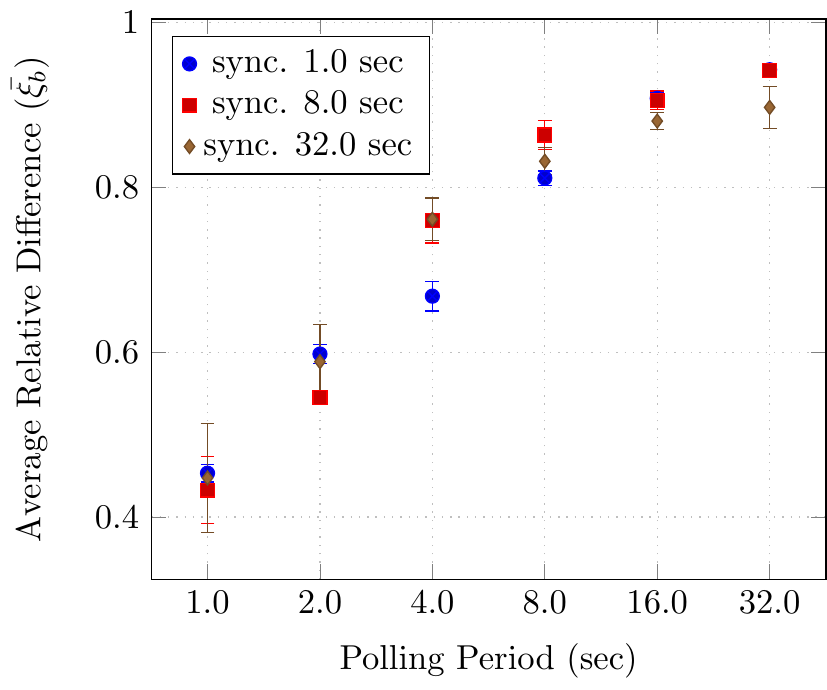}
		\caption{HV}
		\label{fig:D:subfig2}
	\end{subfigure}
}
\caption{Average relative difference ($\bar{\xi_{b}}$) vs polling period in case of a DA LB.}
\label{fig:multi-avgxi-vs-poll}
\end{figure}

\begin{figure}[!t]
\centering
\resizebox{0.5\textwidth}{!}{
	\begin{subfigure}[b]{0.5\textwidth}
		\includegraphics[scale=1.0]{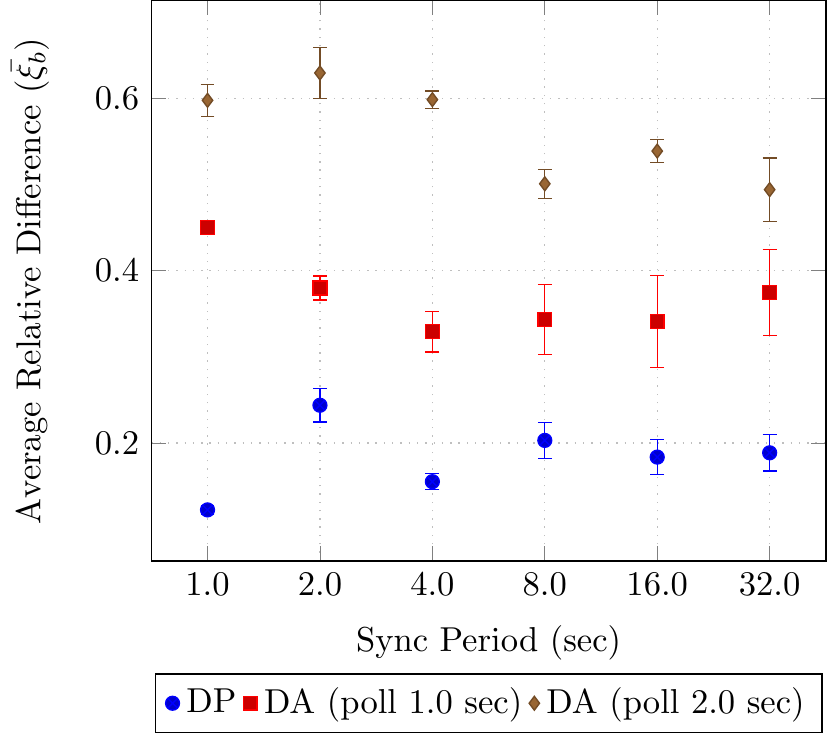}
		\caption{LV}
		\label{fig:E:subfig1}
	\end{subfigure}
	\begin{subfigure}[b]{0.5\textwidth}
		\includegraphics[scale=1.0]{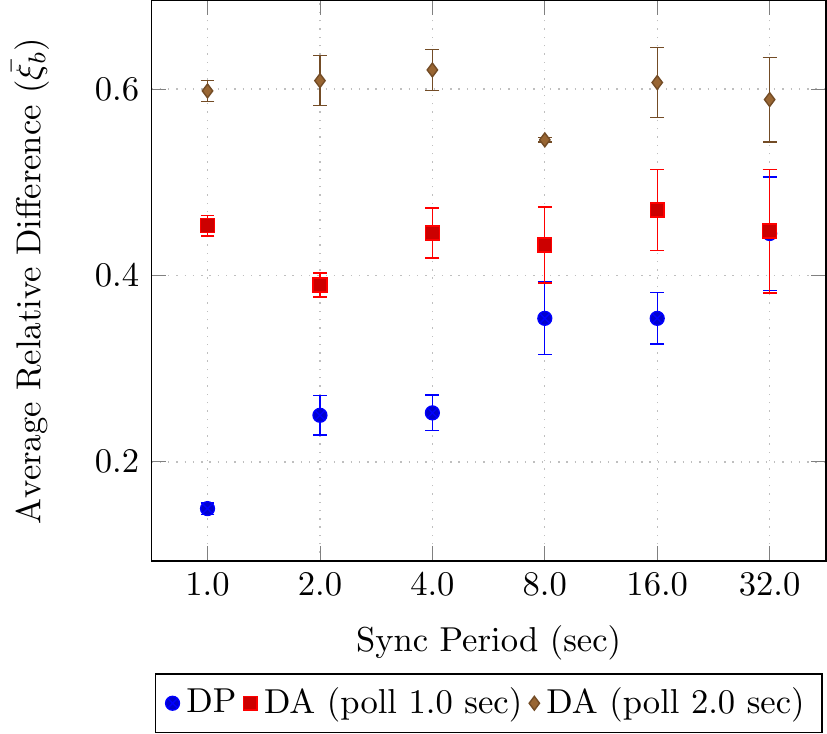}
		\caption{HV}
		\label{fig:E:subfig2}
	\end{subfigure}
}
\caption{Average relative difference ($\bar{\xi_{b}}$) vs synchronization period in case of DP and DA LBs.}
\label{fig:multi-flow-vs-byte}
\end{figure}

\subsection{Results Using Simple Traffic Load}
\label{exp-simple-model}
We  show  the  results  of  our conducted  experiments.
$\bar{\xi}$ represents the relative difference averaged every 2 sec over 10 runs. Recall the definition of the relative difference $\xi_{f}$ (\ref{eqn:xi-flow}) and $\xi_{b}$ (\ref{eqn:xi-byte}).
The smaller the relative difference the better the performance.

Fig. \ref{fig:xi-vs-time} shows how the relative difference in the number of bytes ($\xi_{b}$) measured at each server varies with time (6 sec averaged).
SP performed better most of the time than SA (with polling period 1 and 2 sec).
Comparing Fig. \ref{fig:A:subfig1} and \ref{fig:A:subfig2}, the results show that the traffic load had an impact that was not significant,
\emph{i.e.,} in the LV case the SP performed better than SA even at small polling periods.

Fig. \ref{fig:single-avgxi-vs-poll} shows the effect of the polling period on the performance of the SA LB.
Regardless of the traffic load (Fig. \ref{fig:B:subfig1} and \ref{fig:B:subfig2}), SA is affected by the polling period. As the polling period increases, the performance of SA degrades.

Fig. \ref{fig:multi-avgxi-vs-sync} shows that the performance of DP was affected by the synchronization period among the controllers. As the synchronization period increases, the performance degrades.
Levin et al. \cite{levin2012logically} showed similar results regarding flow-based controllers when measuring the performance in terms of flows.
Relying on flows in decision making (i.e. passive state collection) impacts the results as follows:
(1) the performance of the LB was worse in the case of the HV load than that of the LV load (higher $\bar{\xi_{f}}$ and $\bar{\xi_{b}}$), and
(2) the flow-based $\bar{\xi_{f}}$ and byte-based $\bar{\xi_{b}}$ performance indicators deviated.

Fig. \ref{fig:multi-avgxi-vs-poll} shows the effect of the polling period on the performance of the DA LB, at synchronization periods of 1, 8 and 32 sec.
DA is affected by the polling period. Its performance degrades as the polling period increases, the same as SA (Fig. \ref{fig:single-avgxi-vs-poll}).
Our results show that the polling period has more impact than the synchronization period on DA's performance (same as Fig. \ref{fig:F:subfig1}). As the polling period increases, the state information used by the LB becomes increasingly outdated.
Therefore, for LV traffic load, reducing the frequency of synchronization (i.e. 32 sec) limits the exchange of outdated state information between the controllers.
This is demonstrated in Fig. \ref{fig:D:subfig1} by a lower value of $\bar{\xi_{b}}$ in the case of a 32 sec synchronization period at high polling periods (8, 16 or 32 sec).

Fig. \ref{fig:multi-flow-vs-byte} shows the effect of the synchronization period on the performance of both the DP and the DA LBs.
For LV traffic, DP outperformed DA, even at high synchronization periods (similar to \ref{fig:F:subfig1}). However, for HV traffic the performance of DP started to degrade with the synchronization period.

\subsection{Results Using More Realistic Traffic Load}
\label{exp-realistic-model}
We further experimented with On/Off Pareto \cite{arnold2015pareto} process traffic load. Our On/Off Pareto traffic generator was designed similar to the one included with the NS-2 simulator \cite{ns2pareto}. Equation (\ref{eqn:pareto-pdf}) shows the probability density function of a Pareto distribution, where $\alpha$ is known as the \emph{shape} parameter and $m$ as the \emph{scale} parameter. For a self-similar traffic, $\alpha$ needs to be $<2$; we use a fixed a value of $1.5$ for $\alpha$ (same as the default value in NS-2). Given the mean burst time (also mean On period) and the mean IDLE time (also mean Off period), an appropriate value for $m$ can be calculated as follows: $m_{on} = n_{on} (\frac{\alpha - 1}{\alpha})$, and $m_{off} = \sigma_{off} (\frac{\alpha - 1}{\alpha})$, where $n_{on} = \sigma_{on} * p$. $\sigma_{on}$ is the mean burst time, $\sigma_{off}$ the mean IDLE time and $p$ is packet rate.

\begin{figure}[!t]
\centering
\resizebox{0.5\textwidth}{!}{
	\begin{subfigure}[b]{0.5\textwidth}
		\includegraphics[scale=1.0]{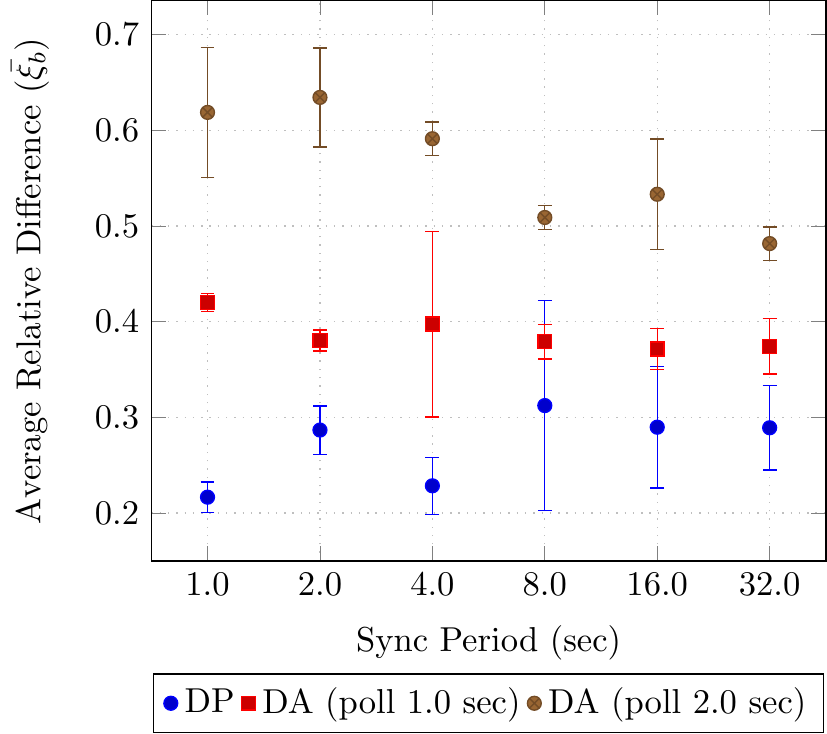}
		\caption{LV}
		\label{paretofig:E:subfig1}
	\end{subfigure}
	\begin{subfigure}[b]{0.5\textwidth}
		\includegraphics[scale=1.0]{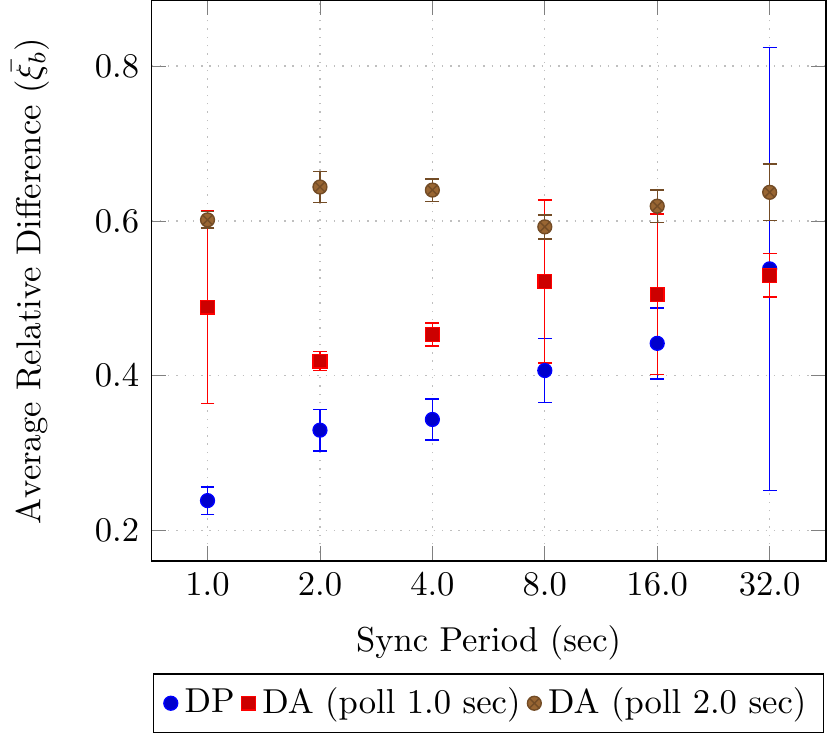}
		\caption{HV}
		\label{paretofig:E:subfig2}
	\end{subfigure}
}
\caption{Average relative difference ($\bar{\xi_{b}}$) vs synchronization period in case of DP and DA LBs using the more realistic work load.}
\label{paretofig:multi-flow-vs-byte}
\end{figure}

Fig. \ref{paretofig:multi-flow-vs-byte} shows the average relative difference ($\xi_{b}$) versus the synchronization period for the cases of DP and DA LBs. The results were obtained using the experimental setup shown in $\S$VIII and with the traffic parameters show in Table \ref{pareto-traffic-param}, they confirm that in the case of DP, the performance of the LB degrades with the increase in synchronization period while the DA was more resilient.

\begin{equation}
	f_{X}(x) =
		\begin{cases}
			\frac{\alpha m ^{\alpha}}{x^{\alpha + 1}} , & x \geq x_{m} \\
			0, & x < x_{m}
		\end{cases}
	\label{eqn:pareto-pdf}
\end{equation}

\begin{table}[h]
	\caption{
	Employed traffic loads and their parameters in case of the more realistic traffic load.
	$r_{1}$ and $r_{2}$ are flow arrival rates for switch 1 and 2, respectively.
	$p_{1}$ and $p_{2}$ are packets-per-flow arrival rates for switch 1 and 2, respectively. Payload of any packet is 4KBytes.
	}
	\centering
	\resizebox{.5\textwidth}{!}{
	\begin{tabular}{l | c c}
	\multicolumn{1}{c}{} & LV & HV\\
	\hhline{-:=:=:}
	Flows & \multicolumn{2}{c}{Poisson process}\\
	Flows rates &  $r_{1}=6 f/s, r{2}=4 f/s$ & $r_{1}=8 f/s, r_{2}=2 f/s$\\
	Flow TTL & \multicolumn{2}{c}{$2 sec$}\\
	\hline
	Pkts-per-flow & \multicolumn{2}{c}{On/Off Pareto \cite{ns2pareto}}\\
	Pkts-per-flow rates & $p_{1}=34 p/s, p_{2}=30 p/s$ & $p_{1}=48 p/s, p_{2}=16 p/s$\\
	\hline
	Pareto \emph{shape} param. & \multicolumn{2}{c}{$1.5$}\\
	Mean burst & $0.6,0.4 sec$  & $0.8,0.2 sec$\\
	Mean idle & $0.4,0.6 sec$ & $0.2,0.8 sec$\\
	\hline
	\end{tabular}
	}
	\label{pareto-traffic-param}
\end{table}

\subsection{Commentary on the Impact of Network State Collection}
Our experimental evaluation shows that in case of LV traffic, where flows are comparable (in byte counts), the application that relied on passive state collection performed better than the one that relied on active state collection. The performance of the application that relied on active state collection was mainly dependent on the polling periods, and in the context of a distributed environment was more affected by the polling periods than the synchronization periods.

Lastly, since the results show that the nature of traffic (LV versus HV) has an impact on the application performance, SDN application developers should pay attention to how they define flows in their applications.

\section{Impact of Network State Collection on State-Distribution-Aware Applications}
\label{sec-lsvs}

Some network application developers could design their applications to be aware of some of the previously mentioned issues in SDN, such as the LSVS-based \cite{Guo201495} LBs, that employ a state distribution mechanism especially designed for distributed LBs. We experimented with two different implementations that we created based on LSVS distributed load-balacers. The first LB (DP-LSVS) uses a passive network state collection mechanism, while the second (DA-LSVS) uses an active network state collection mechanism. The DA-LSVS was tested with polling periods of 1 and 2 sec.

\subsection{Results Using Simple Traffic Load}
\begin{figure}[!t]
\centering
\resizebox{0.5\textwidth}{!}{
	\begin{subfigure}[b]{0.5\textwidth}
		\includegraphics[scale=1.0]{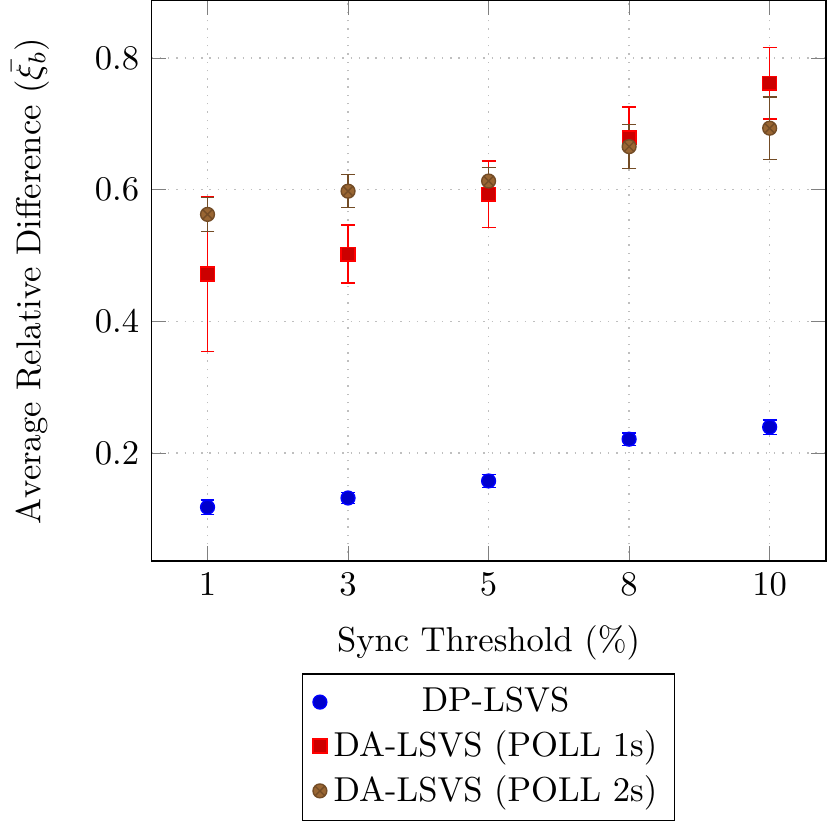}
		\caption{LV}
	\end{subfigure}
	\begin{subfigure}[b]{0.5\textwidth}
		\includegraphics[scale=1.0]{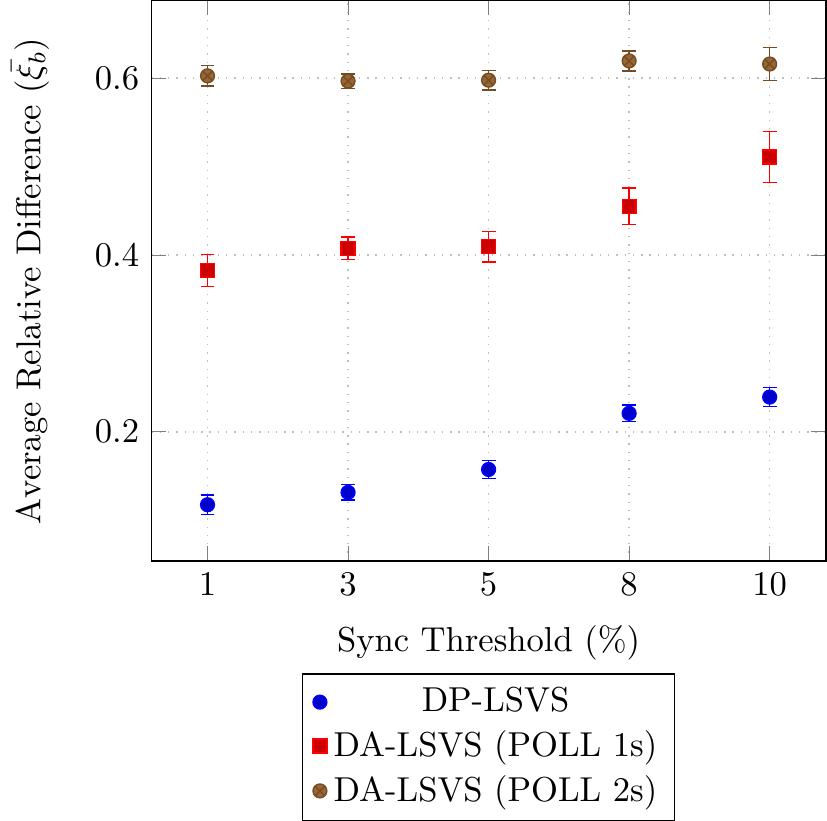}
		\caption{HV}
	\end{subfigure}
}
\caption{Average relative difference ($\bar{\xi_{b}}$) vs synchronization threshold in cases of DP and DA LSVS load-balancers (1 and 2 sec polling periods).}
\label{fig-lsvs}
\end{figure}

In the case of the simple traffic load (shown in Table \ref{traffic-param}), the results shown in Fig. \ref{fig-lsvs} confirm that the synchronization threshold can have an impact on the LB performance. As the threshold increases, the performance of the LB decreases. Nevertheless, the results show that using different network state collection mechanisms can have different impacts on the application performance (for the traffic load we tried, DP-LSVS performed better than DA-LSVS with 1 and 2 sec polling periods). Therefore, network state collection can have an impact on the performance of SDN applications even those that were designed to mitigate the issues of controller state distribution.

\subsection{Results Using More Realistic Traffic Load}
\begin{figure}[!t]
\centering
\resizebox{0.5\textwidth}{!}{
	\begin{subfigure}[b]{0.5\textwidth}
		\includegraphics[scale=1.0]{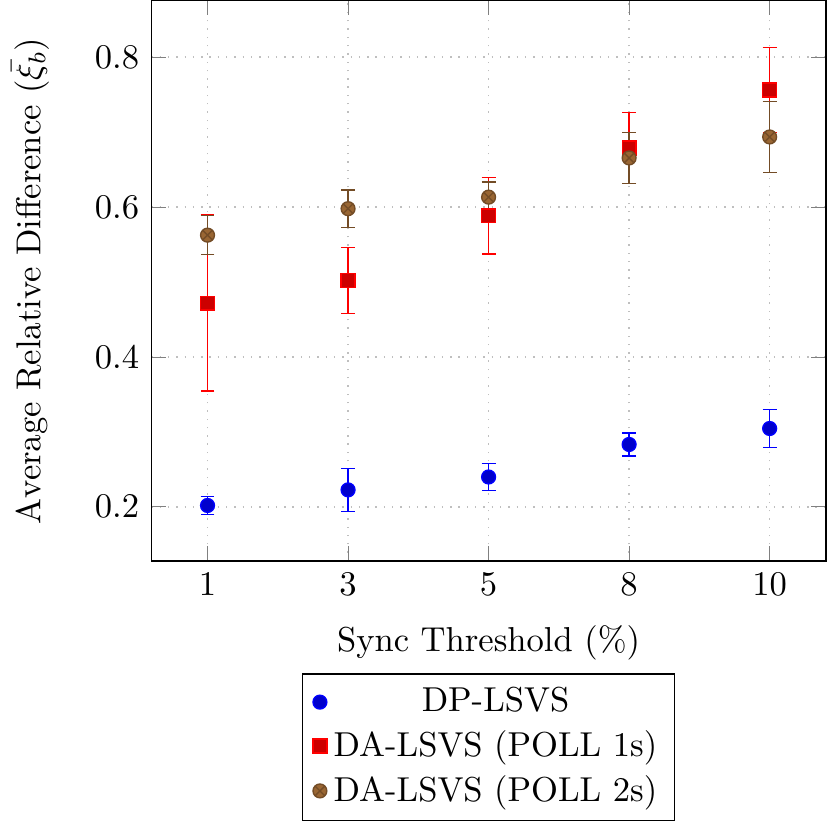}
		\caption{LV}
	\end{subfigure}
	\begin{subfigure}[b]{0.5\textwidth}
		\includegraphics[scale=1.0]{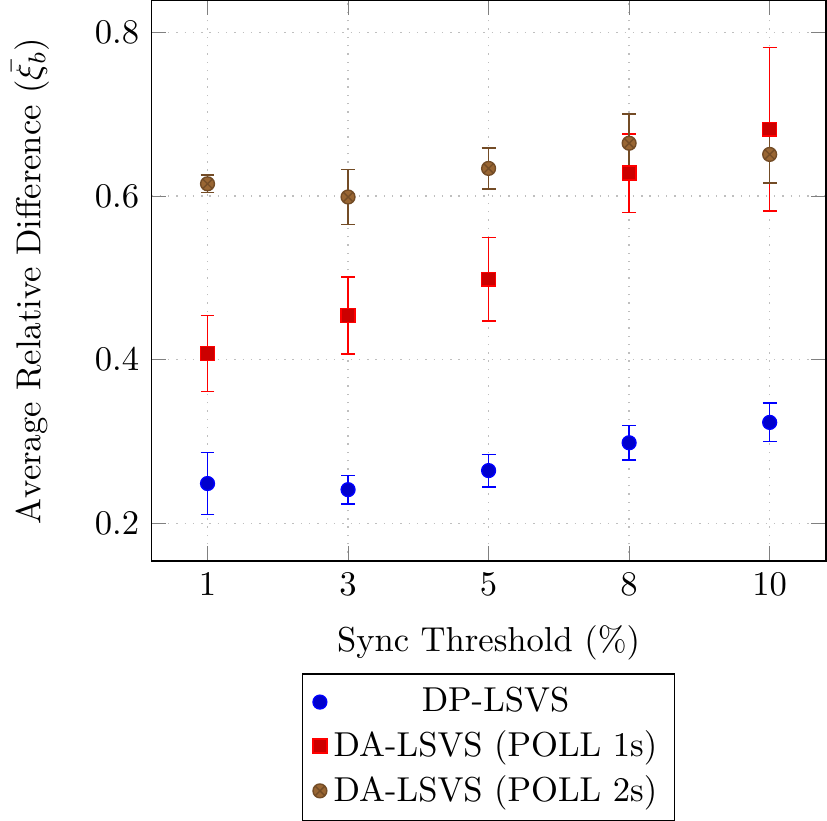}
		\caption{HV}
	\end{subfigure}
}
\caption{Average relative difference ($\bar{\xi_{b}}$) vs synchronization threshold in cases of DP and DA LSVS load-balancers (1 and 2 sec polling periods)  using the more realistic work load.}
\label{paretofig-lsvs}
\end{figure}

Fig. \ref{paretofig-lsvs} shows the average relative difference ($\xi_{b}$) versus the synchronization threshold for the cases of DP and DA LBs, when the more realistic On/Off Pareto traffic load is employed (shown in Table \ref{pareto-traffic-param}). The results confirm the same findings of Fig. \ref{fig-lsvs}.

\section{Model Evaluation Using a Larger Scale Network}
\label{sec-large-scale-net}
As the number of controllers increases, the controller state distribution messages also increase. In this section, we study the impact of network state collection on SDN application performance in case where a higher number of controllers (more than two) was employed. In order to do this, we had to choose a different application performance indicator, as the relative difference ($\xi$) between the amount of traffic received by each server, works only with two controllers. Hence, we opted for the standard deviation ($\sigma$) of the amount of traffic received by each server.

\subsection{The \emph{Standard Deviation} as a Performance Indicator}
We derive the standard deviation of the amount of traffic received by each server ($\sigma$) only in the case of a passive LB (\emph{i.e.,} no polling period). However, similar steps can be used to derive $\sigma$ for active LBs.

We make the following assumptions:
(1) zero synchronization period (\emph{i.e.,} controllers have up-to-date information),
(2) the load-balancing algorithm is invoked for every flow arrival event (as in the case of OpenFlow), and
(3) we do not take network delays into consideration.
Next, we use the following notations:
\begin{itemize}
	\item $N $ --- the number of servers, switches and controllers.
	\item $E^{i}_{k}$ --- the number of expired flows assigned at server $k$ ($1 \le k \le N$) at a given $i^{th}$ inter-arrival event.
	\item $L^{i}_{k}$ --- the load on server $k$, $i$ is the $i^{th}$ flow inter-arrival event. $L^{i}_{k}$ can be in terms of flows (passive LBs) or bytes (active LBs).
	\item $T^{i}$ --- the total load on the servers.
	\begin{equation}
		T^{i} = \sum_{k=1}^{N} L_{k}^{i}
		\label{xload-sum}
	\end{equation}
	\item $\mu^{i}$ --- the average load on the servers.
	\begin{equation}
		\mu^{i} = \frac{T^{i}}{N} = \frac{\sum_{k=1}^{N} L_{k}^{i}}{N}
		\label{xload-avg}
	\end{equation}
	\item $\sigma ^{i}$ --- the standard deviation of the servers' load.
	\begin{equation}
		\sigma^{i} = \sqrt{\frac{1}{N} \sum_{k=1}^{N} (L_{k}^{i} - \mu^{i}) ^ 2}
		\label{xload-std}
	\end{equation}
	\item $M^{i}_{k}$ --- number of new flows that will be assigned to server $k$, based on the second assumption $\sum_{k=1}^{N} M^{i}_{k} = 1$.
	\item $d^{i}_{k}$ --- updated $k^{th}$ server's load as seen by the controller at inter-arrival event $i$.
\end{itemize}
\begin{align}
	d^{i+1}_{k} &= L^{i}_{k} - E^{i+1}_{k} && k \in (0, N]\\
	k^{i+1}_{least} &= \argmin_k d^{i+1}_{k}
\end{align}
\begin{align}
M^{i+1}_{k} =
\begin{cases}
	1, & \text{if } k = d^{i+1}_{min}\\
	0, & otherwise
\end{cases}
\end{align}
\begin{align}
	L^{i+1}_{k} &= L^{i}_{k} - E^{i+1}_{k} + M^{i+1}_{k}
	\label{xnew-load}
\end{align}
\begin{align}
	\mu^{i+1} &= \frac{T^{i+1}}{N} = \frac{\sum_{k=1}^{N} L_{k}^{i+1}}{N}\nonumber\\
				&= \frac{\sum_{k=1}^{N} (L_{k}^{i} - E_{k}^{i+1} + M_{k}^{i+1})}{N} &&(using (\ref{xnew-load}))\nonumber\\
				&= \frac{\sum_{k=1}^{N} L_{k}^{i} - \sum_{k=1}^{N} E_{k}^{i+1} + \sum_{k=1}^{N} M_{k}^{i+1}}{N}\nonumber\\
				&= \mu ^{i} - \frac{\sum_{k=1}^{N} E_{k}^{i+1}}{N} + \frac{1}{N} &&
\end{align}
\begin{align}
	\sigma^{i+1} &= \sqrt{\frac{1}{N} \sum_{k=1}^{N} (L_{k}^{i+1} - \mu^{i+1})^2} &&(using (\ref{xload-std}))\nonumber\\
				&= \sqrt{
						\begin{aligned}				
							\frac{1}{N} \sum_{k=1}^{N} ((L^{i}_{k} - E^{i+1}_{k} + M^{i+1}_{k}) \\
							 - (\mu ^{i} - \frac{\sum_{k=1}^{N} E_{k}^{i+1}}{N} + \frac{1}{N}))^2
						\end{aligned}
				} &&\IEEEQED
\end{align}

In the case of a non-zero synchronization period, the server load values maintained by each controller might differ. For DP controllers, they base their decision on values of $L^{i}$ for the local server and $L^{i-s}$ for  the out-of-domain server, where $s$ is a variable that depends on the synchronization period. For DA controllers, they base their decision on values of $L^{i-p}$ for the local server and $L^{i-s}$ for the out-of-domain server, where $p$ is a variable that depends on the polling period.

\subsection{Results Using Simple Traffic Load}

We show the results of our conducted simulations in case of more than two controllers using the simple traffic load (shown in Table \ref{traffic-param-multi}).
$\bar{\sigma}$ represents the standard deviation averaged every 2 sec (recall (\ref{xload-std})).
$\bar\bar{\sigma}$ represents the average standard deviation ($\bar{\sigma}$) averaged over 10 runs.
The smaller the standard deviation the better the performance.

\begin{figure}[!t]
\centering
\resizebox{0.5\textwidth}{!}{
	\begin{subfigure}[b]{0.5\textwidth}
		\includegraphics[scale=0.8]{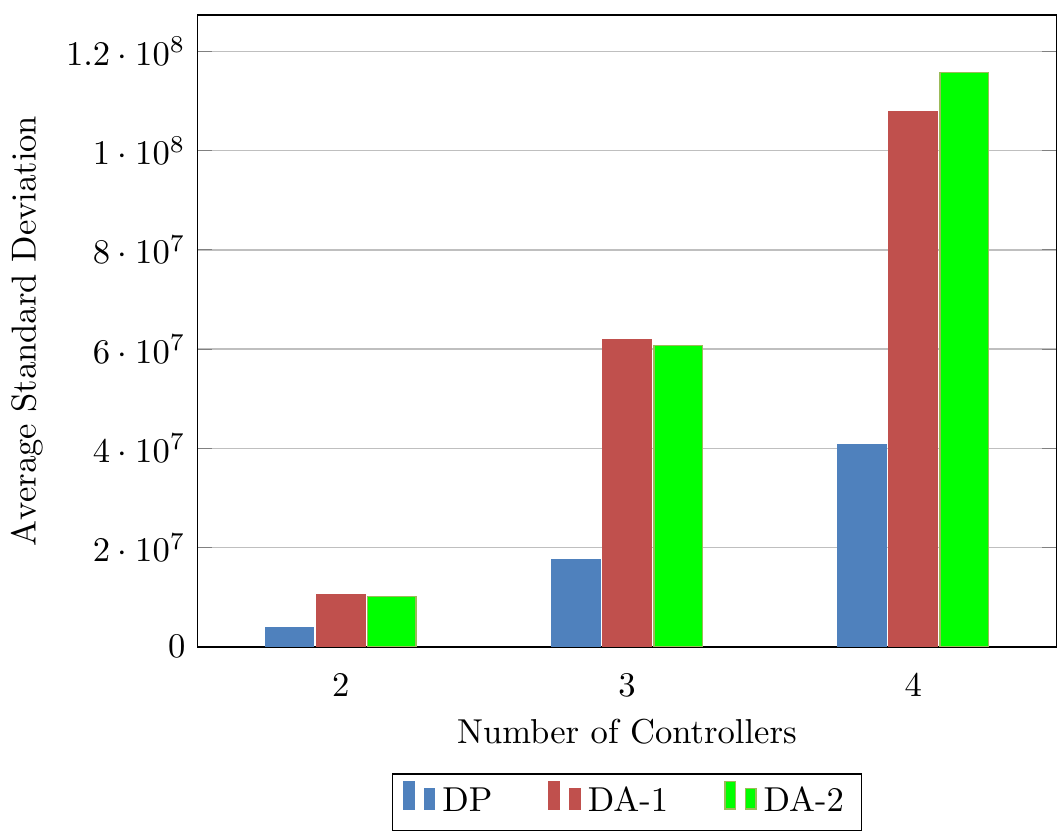}
		\caption{LV}
		\label{fig:H:subfig1}
	\end{subfigure}
	\begin{subfigure}[b]{0.5\textwidth}
		\includegraphics[scale=0.8]{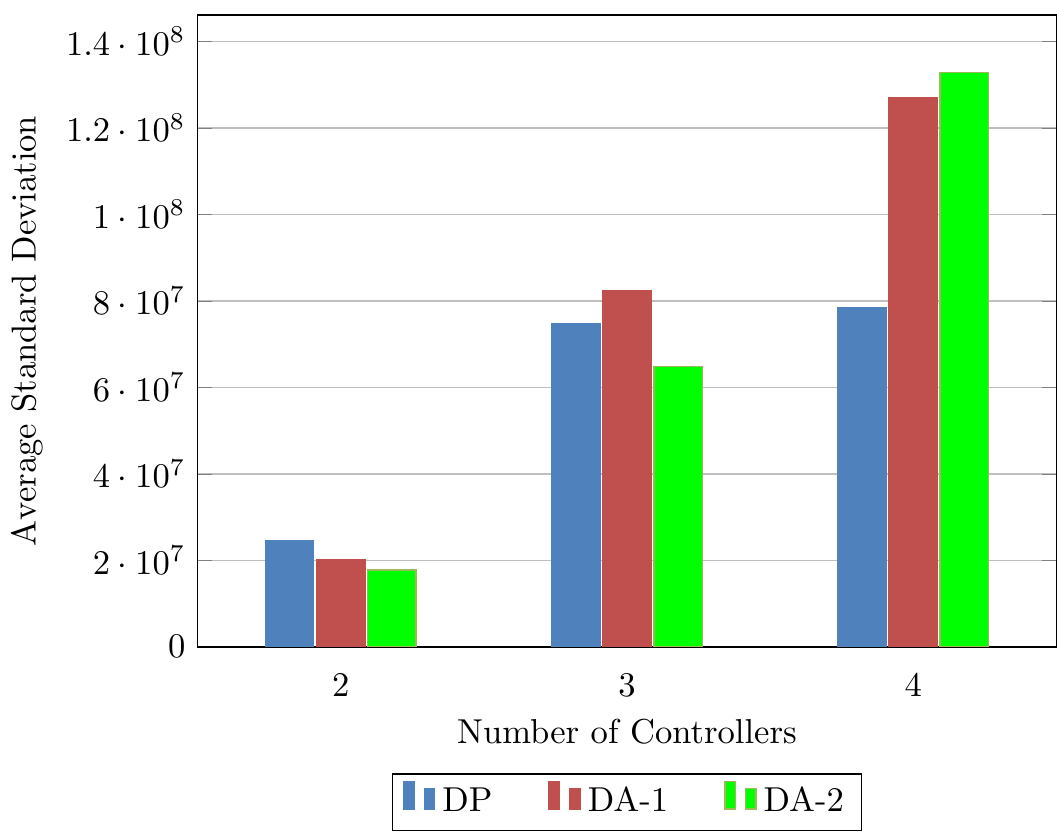}
		\caption{HV}
		\label{fig:H:subfig2}
	\end{subfigure}
}
\caption{Average standard deviation for all sync periods ($\bar{\bar{\sigma_{b}}}$) vs number of controllers using simple traffic load}
\label{fig:mu-vs-controllers}
\end{figure}

Fig. \ref{fig:mu-vs-controllers} shows how the standard deviation in the number of bytes ($\sigma_{b}$) measured at each server varies with the number of controllers in both cases of DP and DA (with polling period 1 and 2 sec) LBs.
The results show that the impact of network state collection on SDN application performance increases with the increase of the number of distributed controllers employed.

\begin{table}[h]
	\caption{
	Employed traffic loads and their parameters in case of the simple traffic load.
	$C$ is the number of distributed controllers.
	$r_{lv}$ and $r_{hv}$ are the LV and HV flow arrival rates for the switches, respectively.
	$p_{lv}$ and $p_{hv}$ are the LV and HV packets-per-flow arrival rates for the switches, respectively. Payload of any packet is 4KBytes.
	}
	\centering
	\resizebox{.5\textwidth}{!}{
	\begin{tabular}{l | c | c | c}
	\multicolumn{1}{c}{} & $C=2$ & $C=3$ & $C=4$\\
	\hhline{-:=:=:=:}
	Flows & \multicolumn{3}{c}{Poisson process}\\
	\multirow{2}{*}{Flows rates} & $r_{lv} = \{6, 4\} f/s$ & $r_{lv} = \{6, 4, 5\} f/s$ & $r_{lv} = \{6, 4, 5, 4\} f/s$\\
								& $r_{hv} = \{8, 2\} f/s$ & $r_{hv} = \{8, 2, 4\} f/s$ & $r_{hv} = \{8, 2, 4, 6\} f/s$\\
	Flow TTL & \multicolumn{3}{c}{$2 sec$}\\
	Pkts-per-flow & \multicolumn{3}{c}{Poisson process}\\
	\multirow{2}{*}{Pkts-per-flow rates} & $p_{lv} = \{34, 30\} p/s$ & $p_{lv} = \{34, 30, 32\} p/s$ & $p_{lv} = \{34, 30, 32, 28\} p/s$\\
										& $p_{hv} = \{48, 16\} p/s$ & $p_{hv} = \{48, 16, 32\} p/s$ & $p_{hv} = \{48, 16, 32, 20\} p/s$\\
	\hline
	\end{tabular}
	}
	\label{traffic-param-multi}
\end{table}

\subsection{Results Using More Realistic Traffic Load}
\begin{figure}[!t]
\centering
\resizebox{0.5\textwidth}{!}{
	\begin{subfigure}[b]{0.5\textwidth}
		\includegraphics[scale=0.8]{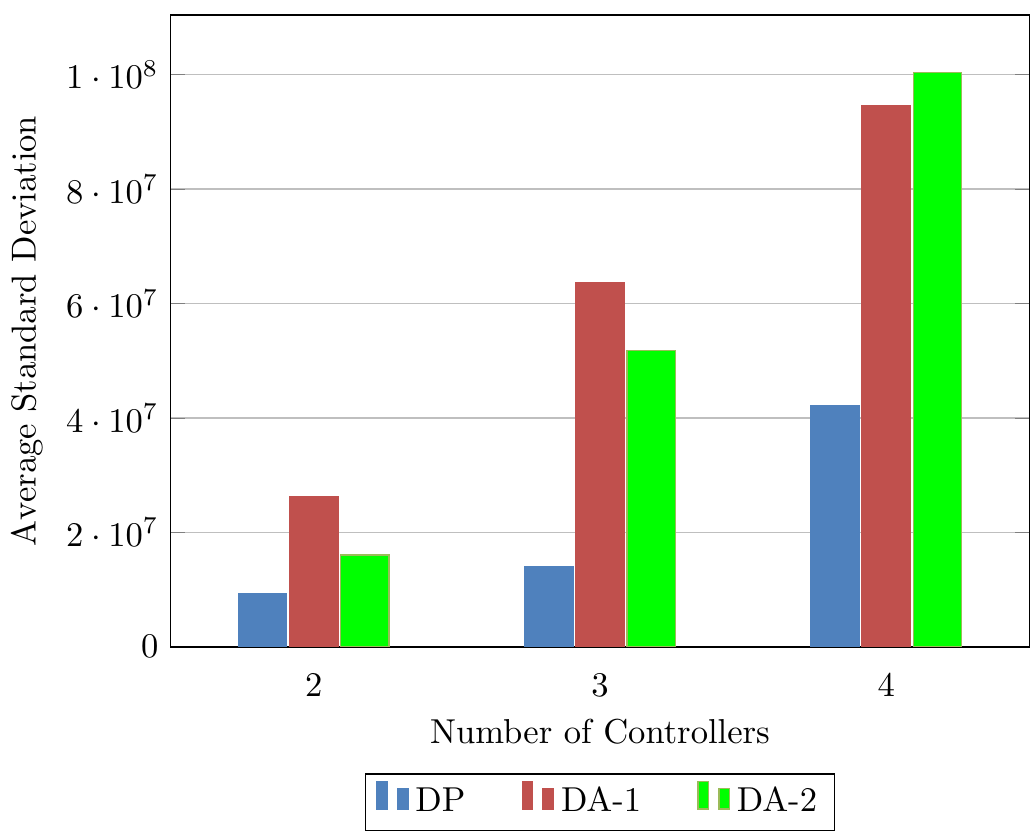}
		\caption{LV}
		\label{paretofig:H:subfig1}
	\end{subfigure}
	\begin{subfigure}[b]{0.5\textwidth}
		\includegraphics[scale=0.8]{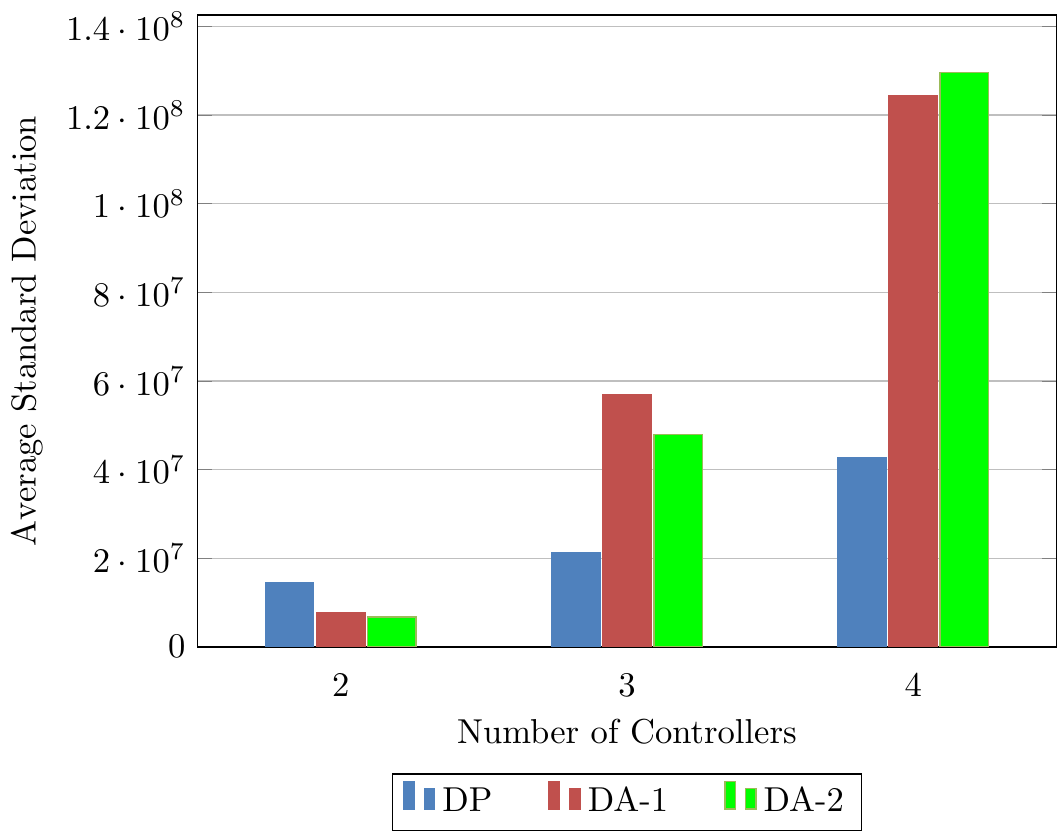}
		\caption{HV}
		\label{paretofig:H:subfig2}
	\end{subfigure}
}
\caption{Average standard deviation for all sync periods ($\bar{\bar{\sigma_{b}}}$) vs number of controllers using more realistic traffic load.}
\label{paretofig:mu-vs-controllers}
\end{figure}

We  present  the  results  of our experiments in case of more than two controllers using the more realistic On/Off Pareto traffic load (shown in Table \ref{pareto-traffic-param}).
Fig. \ref{paretofig:mu-vs-controllers} shows how the standard deviation in the number of bytes ($\sigma_{b}$) measured at each server varies with the number of controllers in both cases of DP and DA (with polling period 1 and 2 sec) LBs.
The results again confirm that the impact of network state collection on SDN application performance increases with the increase of the number of distributed controllers employed.

\begin{table}[h]
	\caption{
	Employed traffic loads and their parameters in case of more realistic traffic load.
	$C$ is the number of distributed controllers.
	$r_{lv}$ and $r_{hv}$ are the LV and HV flow arrival rates for the switches, respectively.
	$p_{lv}$ and $p_{hv}$ are the LV and HV packets-per-flow arrival rates for the switches, respectively.
	Payload of any packet is 4KBytes.
	}
	\centering
	\resizebox{.5\textwidth}{!}{
	\begin{tabular}{l | c | c | c}
	\multicolumn{1}{c}{} & $C=2$ & $C=3$ & $C=4$\\
	\hhline{-:=:=:=:}
	Flows & \multicolumn{3}{c}{Poisson process}\\
	\multirow{2}{*}{Flows rates} & $r_{lv} = \{6, 4\} f/s$ & $r_{lv} = \{6, 4, 5\} f/s$ & $r_{lv} = \{6, 4, 5, 4\} f/s$\\
								& $r_{hv} = \{8, 2\} f/s$ & $r_{hv} = \{8, 2, 4\} f/s$ & $r_{hv} = \{8, 2, 4, 6\} f/s$\\
	Flow TTL & \multicolumn{3}{c}{$2 sec$}\\
	Pkts-per-flow & \multicolumn{3}{c}{On/Off Pareto \cite{ns2pareto}}\\
	\multirow{2}{*}{Pkts-per-flow rates} & $p_{lv} = \{34, 30\} p/s$ & $p_{lv} = \{34, 30, 32\} p/s$ & $p_{lv} = \{34, 30, 32, 28\} p/s$\\
										& $p_{hv} = \{48, 16\} p/s$ & $p_{hv} = \{48, 16, 32\} p/s$ & $p_{hv} = \{48, 16, 32, 20\} p/s$\\
	Pareto \emph{shape} param. & \multicolumn{3}{c}{$1.5$}\\
	Mean burst & $0.6,0.4 sec$  & $0.6,0.4,0.5 sec$ & $0.6,0.4,0.5,0.8 sec$\\
	Mean idle & $0.4,0.6 sec$ & $0.4,0.6,0.5 sec$ & $0.4,0.6,0.5,0.2 sec$\\
	\hline
	\end{tabular}
	}
	\label{pareto-traffic-param-multi}
\end{table}

\section{Conclusion and Future Work}
\label{sec-conclusion}

In this paper, we studied two key factors affecting the global network view at the SDN controllers, particularly network state collection and controller state distribution.
We compared the impact of active and passive OpenFlow network state collection methods on an SDN load-balancing application in the context of both single and distributed controller environments.
Our evaluation showed that in case of low-variation traffic, where flows are comparable (in byte counts), the application that relied on passive state collection performed better than the one that relied on active state collection. The performance of the application that relied on active state collection was mainly dependent on the polling periods, and in the context of a distributed environment was more affected by the polling periods than the synchronization periods.
Even with LBs that were designed to overcome the controllers state distribution problem such as LSVS-based LBs, they were still vulnerable to the impact of network state collection.
Since the results showed that the nature of traffic (LV versus HV) has an impact on the application performance, SDN application developers should pay attention to how they define flows in their applications.
Lastly, the results showed that the impact of network state collection on the SDN application performance becomes more apparent (performance decreases) with the increase in the number of controllers (network scale).

For future work, we plan to investigate the impact of network state collection and state distribution on the performance of other single and distributed SDN applications (e.g. security applications) using different performance indicators. We would further like to conduct our experiments on a larger scale network.

In \cite{aslan2016adaptive}, we attempted to mitigate the impact of controller state distribution on SDN application performance by suggesting the use of adaptive controllers into SDN.

\begin{table}[]
\centering
\caption{Summary of Conducted Experiments}
\label{tbl-exp-summary}
\begin{tabular}{l|l|c|c|c|c}
\cline{3-6}
\multicolumn{2}{l}{\multirow{2}{*}{}}   & $\xi$ model   		& $\xi$ exp. 								& $\sigma$ model & LSVS \\ \cline{3-6} 
\multicolumn{2}{l}{}                    & Simulation		& Evaluation    										& Simulation      & Evaluation \\ \hline
\multirow{4}{*}{LB}       & SA            &         		& Fig. \ref{fig:xi-vs-time}, \ref{fig:single-avgxi-vs-poll}       		&          &      \\ 
                          & SP            &         		& Fig. \ref{fig:xi-vs-time}    					    		&          &      \\ 
                          & DA            & Fig. \ref{fig:math-multi-flow-vs-byte}        		& Fig. \ref{fig:multi-avgxi-vs-poll}, \ref{fig:multi-flow-vs-byte}, \ref{paretofig:multi-flow-vs-byte}      		& Fig. \ref{fig:mu-vs-controllers}, \ref{paretofig:mu-vs-controllers}         & Fig. \ref{fig-lsvs}, \ref{paretofig-lsvs}      \\ 
                          & DP            & Fig. \ref{fig:math-multi-flow-vs-byte}       		& Fig. \ref{fig:multi-avgxi-vs-sync}, \ref{fig:multi-flow-vs-byte}, \ref{paretofig:multi-flow-vs-byte}									& Fig. \ref{fig:mu-vs-controllers}, \ref{paretofig:mu-vs-controllers}          & Fig. \ref{fig-lsvs}, \ref{paretofig-lsvs}     \\ \hline
\multirow{5}{*}{Perf. vs} & Time          &       			& Fig. \ref{fig:xi-vs-time}      		 												    &          &      \\ 
                          & Sync          & Fig. \ref{fig:math-multi-flow-vs-byte}    		    & Fig. \ref{fig:multi-avgxi-vs-sync}, \ref{fig:multi-flow-vs-byte}, \ref{paretofig:multi-flow-vs-byte}     		 												    &          &      \\ 
                          & Poll          &   			    & Fig. \ref{fig:single-avgxi-vs-poll}, \ref{fig:multi-avgxi-vs-poll}											&          &      \\ 
                          & Thres.	   &    		    &      														    &		& Fig. \ref{fig-lsvs}, \ref{paretofig-lsvs}     \\ 
                          & \# Cntrl &    		    &      														    & Fig. \ref{fig:mu-vs-controllers}, \ref{paretofig:mu-vs-controllers}          &      \\ \hline
                          
\end{tabular}
\end{table}

\section*{Acknowledgment}
The second author acknowledge support from the Natural Sciences and Engineering Research Council of Canada (NSERC) through the NSERC Discovery Grant program.

\ifCLASSOPTIONcaptionsoff
  \newpage
\fi

\bibliographystyle{IEEEtran}
\bibliography{references}

\begin{IEEEbiography}[{\includegraphics[width=1in,height=1.25in,clip,keepaspectratio]{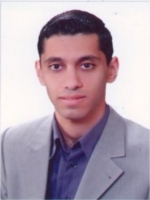}}]{Mohamed Aslan}
is a PhD candidate in the Department of Systems and Computer Engineering at Carleton University. In 2012, he received his M.Sc. degree in Computer Engineering from the Arab Academy for Science and Technology. His research is mainly in the areas of future internet architectures and distributed systems, with current focus on software-defined networks.
\end{IEEEbiography}

\begin{IEEEbiography}[{\includegraphics[width=1in,height=1.25in,clip,keepaspectratio]{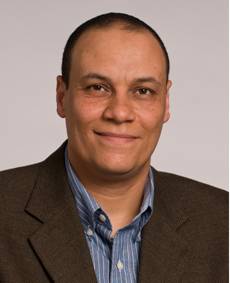}}]{Ashraf Matrawy}
is an Associate Professor of the School of Information Technology at Carleton University. He is a senior member of the IEEE and serves on the editorial board of the IEEE Communications Surveys and Tutorials journal. He has served as a technical program committee member of IEEE CNS, IEEE ICC, IEEE Globecom, IEEE LCN, and IEEE/ACM CC-GRID. He is also a Network co-Investigator of Smart Cybersecurity Network (SERENE-RISC). His research interests include reliable and secure computer networking, software-defined networking and cloud computing.
\end{IEEEbiography}

\end{document}